\documentclass[runningheads]{llncs}
\usepackage[T1]{fontenc}
\usepackage{graphicx}
\usepackage{subcaption}
\usepackage{algorithmic}
\usepackage{upgreek}
\usepackage{multirow}
\usepackage{amsmath}
\usepackage{float}
\usepackage{hyperref}
\begin{document}
\title{Predicting Satisfaction of Counterfactual Explanations from Human Ratings of Explanatory Qualities}
%
%
\author{Marharyta Domnich\inst{1}\orcidID{0000-0001-5414-6089} \and
Rasmus Moorits Veski\inst{1,2}, Julius Välja\inst{1}, Kadi Tulver\inst{1}\orcidID{0000-0002-6263-4098} \and
Raul Vicente\inst{1}\orcidID{0000-0002-2497-0007}}
\authorrunning{M. Domnich et al.}
%
\institute{Institute of Computer Science, University of Tartu, Tartu, Estonia \and
École Polytechnique Fédérale de Lausanne, Lausanne, Switzerland
\email{marharyta.domnich@ut.ee}}
\maketitle              
\begin{abstract}
Counterfactual explanations are a widely used approach in Explainable AI, offering actionable insights into decision-making by illustrating how small changes to input data can lead to different outcomes. Despite their importance, evaluating the quality of counterfactual explanations remains an open problem. Traditional quantitative metrics, such as sparsity or proximity, fail to fully account for human preferences in explanations, while user studies are insightful but not scalable. Moreover, relying only on a single overall satisfaction rating does not lead to a nuanced understanding of why certain explanations are effective or not. 
To address this, we analyze a dataset of counterfactual explanations that were evaluated by 206 human participants, who rated not only overall satisfaction but also seven explanatory criteria: feasibility, coherence, complexity, understandability, completeness, fairness, and trust. Modeling overall satisfaction as a function of these criteria, we find that feasibility (the actionability of suggested changes) and trust (the belief that the changes would lead to the desired outcome) consistently stand out as the strongest predictors of user satisfaction, though completeness also emerges as a meaningful contributor. Crucially, even excluding feasibility and trust, other metrics explain 58\% of the variance, highlighting the importance of additional explanatory qualities. Complexity appears independent, suggesting more detailed explanations do not necessarily reduce satisfaction. 
Strong metric correlations imply a latent structure in how users judge quality, and demographic background (e.g., medical or ML expertise) significantly affects ranking patterns, highlighting the need for context‐specific designs. These insights directly inform the development of improved counterfactual algorithms, highlighting the need to tailor explanatory qualities (completeness, consistency, fairness, complexity) to diverse user expertise and specific domain contexts.

\keywords{Counterfactual Explanations  \and Explainable AI \and Human-Centric Evaluation \and Explanatory Virtues.}
\end{abstract}
\section{Introduction}

The increasing scale and complexity of AI models has spurred development of post-hoc explainability techniques in Explainable AI (XAI). Among these, counterfactual explanations have rapidly emerged as a practical method to show how minimal alterations to an input can change the model’s output \cite{wachter2017counterfactual}. For instance, Keane’s famous example \cite{keane2021bettercounterfactualexplanationskey} illustrates an automated paper review system stating, “if your paper had more novelty, it would have been accepted to this conference”. Such explanations are inherently contrastive and actionable, guiding users towards potential modifications to achieve a desired result. Furthermore, they align closely with human reasoning, where counterfactual alternatives are naturally considered \cite{miller_explanation_2019}.

Despite the promise of counterfactual explanations, evaluating their effectiveness for users remains an open challenge. Traditional XAI evaluations have relied heavily on proxy metrics and intuition rather than direct user feedback~\cite{keane2021bettercounterfactualexplanationskey}. Automated metrics, such as validity (whether the counterfactual flips the model’s prediction), proximity (the degree of change from the original input), sparsity (the number of altered features), plausibility or data fidelity (the realism of the counterfactual), and diversity (the availability of distinct alternatives)~\cite{guidotti_counterfactual_2022,karimi2022survey}, are useful for algorithmic comparisons but “often fall short in capturing the human perspective” \cite{domnich2024towards}.
An explanation that scores well on proximity and sparsity, for example, might still confuse a user or leave out information the user considers important as “excellent computational explanations may not be good psychological explanations.” \cite{keane2021bettercounterfactualexplanationskey}. 

User studies, which are considered the gold standard for evaluating explanation quality, remain surprisingly scarce in the XAI literature. Keane (2021) reported that only about 21\% of 100 counterfactual explanation studies incorporated human evaluations, and Adadi and Berrada (2018) stated that the “neglect of user studies is the original sin” of the field \cite{adadi2018peeking}.

Direct user feedback can reveal qualities like clarity or usefulness that automated metrics overlook. However, user studies are costly and difficult to scale: recruiting a large number of participants (often domain experts) for evaluating explanations is time-consuming and expensive. Additionally, when user studies are conducted, their findings are difficult to generalize, as user preferences and mental models vary widely and what counts as a "good" explanation can differ from person to person. Recently, researchers have begun exploring large language models (LLMs) to mimic human evaluations as a way to scale these studies \cite{domnich2024towards,de2024evaluating}. Although offering valuable complementary insights, LLM-based assessments cannot yet fully capture the nuanced sentiment provided by direct human feedback. 

However, simply asking users to provide an overall satisfaction score makes it difficult to pinpoint which underlying factors drive that satisfaction, where \textit{understanding what contributes to users' overall satisfaction and how to model it through other measurable explanatory qualities is the main direction of this paper.}
In our prior study, we developed a benchmark dataset of 30 diverse counterfactual scenarios (spanning different domains and qualities), each evaluated by 206 human participants on eight evaluation metrics \cite{domnich_2024_14672264}. These metrics included Overall Satisfaction as a summary measure, alongside specific explanatory qualities such as Feasibility, Consistency, Completeness, Trust, Fairness, Complexity, and Understandability. In this study, we extend that line of research by shifting the focus to modeling overall user satisfaction based on the underlying explanatory metrics. Our goal is to understand and predict how satisfied a user will be with a counterfactual explanation given its scores on various quality dimensions. We explicitly investigate the relationship between the individual metric ratings and the overall satisfaction judgment. To guide this investigation, we pose the following research questions:
\begin{itemize}
    \item Can overall satisfaction be accurately predicted using these seven evaluation metrics (Feasibility, Trust, Consistency, Completeness, etc.)?
    \item Among the chosen metrics, which ones are truly essential for predicting overall satisfaction, and to what extent can we simplify (i.e., omit certain metrics) while retaining high predictive accuracy?
\end{itemize}

\section{Related Works}

Counterfactual explanations resonate with natural human cognitive processes, as individuals often engage in "what if" scenarios to understand events around them. Cognitive science research indicates that people often consider alternatives to reality to understand causal relationships or to learn from past actions. Byrne’s studies, for example, highlight that humans process counterfactual conditionals differently from direct causal statements, which can lead to unique inferences and sometimes suppress certain logical outcomes \cite{byrne2005rational}.
While counterfactual explanations align with natural human reasoning patterns, making them intuitively appealing, they can also introduce cognitive complexities that may not align with the designers' intentions. In later works, Byrne argues that the interpretation of causal versus counterfactual explanations can lead to fundamentally different mental models of the same situation, sometimes enhancing understanding but not necessarily increasing satisfaction \cite{Byrne2019}.
For example, Warren et al. observed that while counterfactual explanations provided users with a richer mental model of decision-making processes in AI systems, they were not always preferred over simpler explanations \cite{warren2023categorical}. \textit{This discrepancy highlights a critical gap between what enhances performance or understanding and what users find most satisfying or intuitive.}

Recent studies further report mixed effects: while counterfactual explanations can enhance objective comprehension, they do not always boost subjective satisfaction \cite{vannostrand2024actionable}

For example, Van der Waa et al. observed that although counterfactual explanations led to higher satisfaction and trust, they did not significantly improve prediction accuracy compared to simpler approaches \cite{van2021interpretable}. The opposite effect is demonstrated in Wang and Yin's user study, where counterfactual explanations improved users’ factual comprehension of AI decisions (measured by prediction accuracy on test cases), but did not reliably increase users’ subjective confidence or trust in the model \cite{wang2021explanations}.
This reflects a gap between performance and preference: an explanation type that best supports accuracy of understanding may not be the one that users find most satisfying or intuitive.

Another insight from cognitive science is that humans do not always favor the simplest explanation. While Occam’s razor holds in science, people sometimes equate a “good” explanation with a more comprehensive account of causes. Zemla et al., for instance, found that for everyday phenomena, people preferred an explanation that combined multiple contributing factors (e.g. policy, demographics, and behavior in explaining a social trend) over an explanation that cites a single cause \cite{zemla2017evaluating}. Participants tended to feel that an event was not fully explained until all salient causes were mentioned, even if the added causes were not actually factual. This suggests that a minimal counterfactual (changing just one factor) might appear to users as too shallow. The effect is more evident in specialized domains like medicine, where medical doctors preferred explanations with the biggest number of causes \cite{barbu2024exploring}. In contrast, XAI algorithmic approaches often prioritize sparsity (altering only one or few features) to generate counterfactuals \cite{mothilal_explaining_2020,codice_domnich_2024}. \textit{Therefore, explanations that make sense to humans may require going beyond algorithmic intuition, ensuring that the format and the content of counterfactual explanations truly support human decisions. }

Supporting this perspective, VanNostrand noted that while users appreciated understanding the decision-making process, their overall satisfaction was heavily influenced by the explanation’s alignment with an intuitive sense of fairness and their perceived ability to act on the provided information \cite{vannostrand2024actionable}.
Similarly, Förster et al. reported that coherent counterfactual explanations were associated with higher overall satisfaction \cite{forster2021capturing}.
\textit{This highlights that user overall satisfaction of explanation is more complex than mere comprehension of the model's operation. }
In our prior work \cite{domnich2024towards}, we compiled a list of cognitive biases informed by the literature and observed a positive correlation with satisfaction for seven evaluation metrics: feasibility, consistency, completeness, trust, understandability, fairness, and complexity. Building on these insights, our current work aims to analyze how Overall Satisfaction can be predicted from these underlying explanatory qualities. Notably, we have not identified any studies that analyze the intercomparison of these metrics in predicting overall satisfaction, which highlights a gap in the existing literature.

\section{Human Evaluation of Counterfactual Explanations}
We base our study on the CounterEval dataset \cite{domnich_2024_14672264}, a recent human evaluation benchmark for counterfactual explanation evaluation. This dataset consists of 30 diverse counterfactual scenarios evaluated by 206 human respondents across 8 key metrics: Overall Satisfaction, Feasibility, Consistency, Completeness, Trust, Fairness, Complexity, and Understandability \cite{domnich_2024_14672264}, resulting in a total of 6180 individual ratings. Each participant rated each scenario on these metrics, providing a data cube of scores (participants $\times$ scenarios $\times$ metrics). All metrics are measured on a Likert-type scale (from 1 to 6 for all metrics, except Complexity which is measured from -2 to 2), allowing comparative quantitative analysis. The scenarios are based on actual outputs from counterfactual frameworks on tabular datasets like Adult and the Pima Indian Diabetes, commonly used in counterfactual explanation algorithms evaluations \cite{rasouli2024care}. In some cases, the outputs were modified to intentionally span a wider range in explanatory quality metrics, while in others, features were tailored to improve clarity for human evaluators after pilot study (more details in \cite{domnich_2024_14672264}). Based on exclusion criteria, 10 participants were removed who failed on three or more of these criteria: attention check, response time, average understandability score, response clustering and three indicator questions.

\subsection{Exploratory Analysis}
Using this dataset, our methodology combines statistical analysis and machine learning to identify which explanation metrics drive overall satisfaction and to build predictive models of satisfaction. 

\begin{figure}[!htbp]
\includegraphics[width=\textwidth]{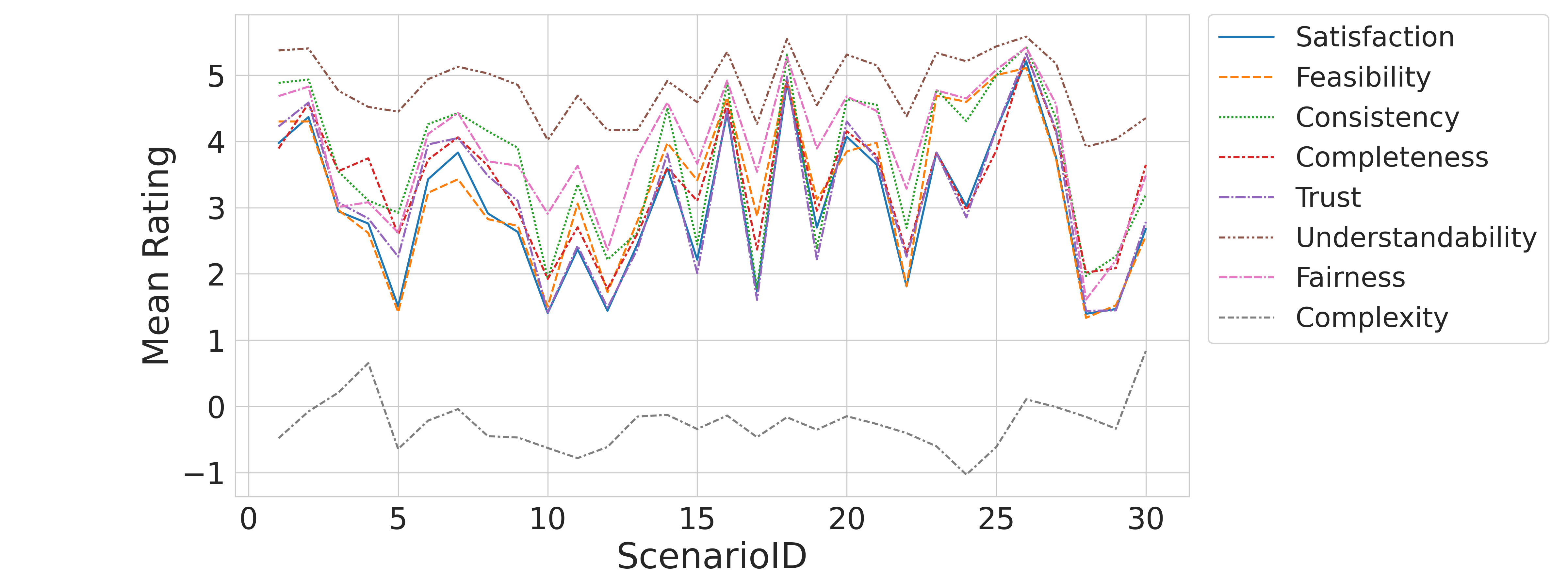}
\caption{Mean ratings of each metric (Overall Satisfaction, Feasibility, Consistency, Completeness, Trust, Understandability, Fairness, and Complexity) across all 30 scenarios. The y-axis shows the average participant rating per scenario, with Complexity on a -2 to +2 scale (0 = ideal complexity) and the other metrics on a 1–6 scale.} \label{mean_metric_scenario}
\end{figure}

Previous work \cite{domnich2024towards} reported strong correlations for all metrics. To investigate this further, the line plot (Figure~\ref{mean_metric_scenario}) shows the average rating of each metric for each of the 30 scenarios. Most metrics track relatively close together, indicating that scenarios viewed favorably (or unfavorably) on one metric often receive similar evaluations on others. Complexity (gray dashed line) follows a separate -2 to +2 range, showing that some scenarios were considered slightly "too simple" or "too complex," while many scenarios clustered near the "desired" complexity level 0. 
\begin{figure}[!htbp]
    \centering
    
    \begin{subfigure}[t]{0.43\textwidth}
        \centering
        \includegraphics[width=\textwidth]{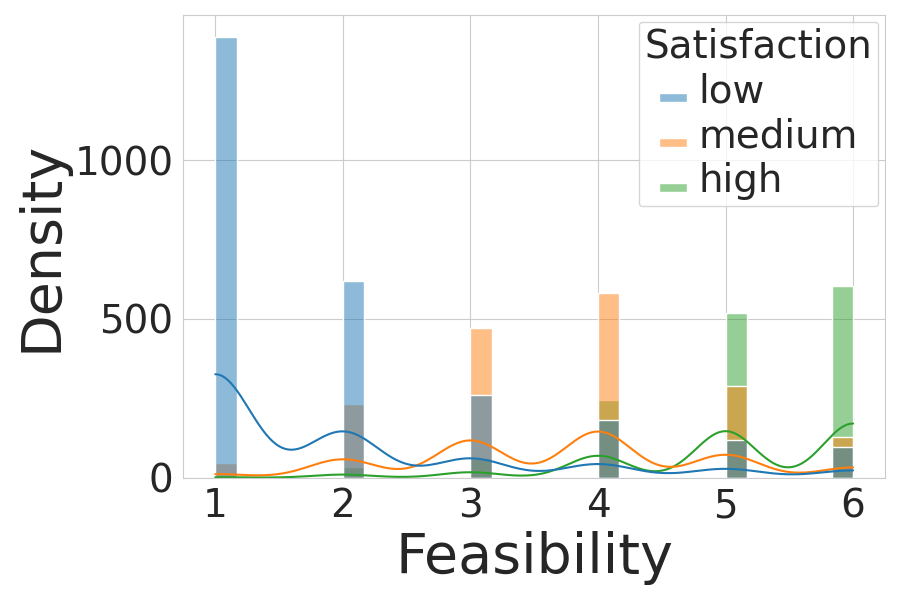}
        \caption*{}
        \label{fig:feasibility_dist}
    \end{subfigure}
    \hfill
    \begin{subfigure}[t]{0.43\textwidth}
        \centering
        \includegraphics[width=\textwidth]{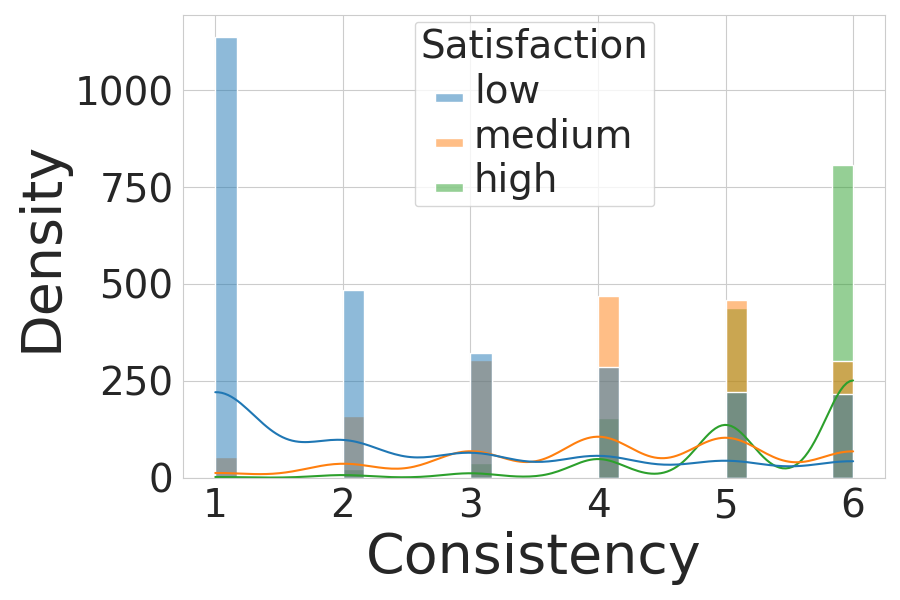}
        \caption*{}
        \label{fig:consistency_dist}
    \end{subfigure}

    \begin{subfigure}[t]{0.43\textwidth}
        \centering
        \includegraphics[width=\textwidth]{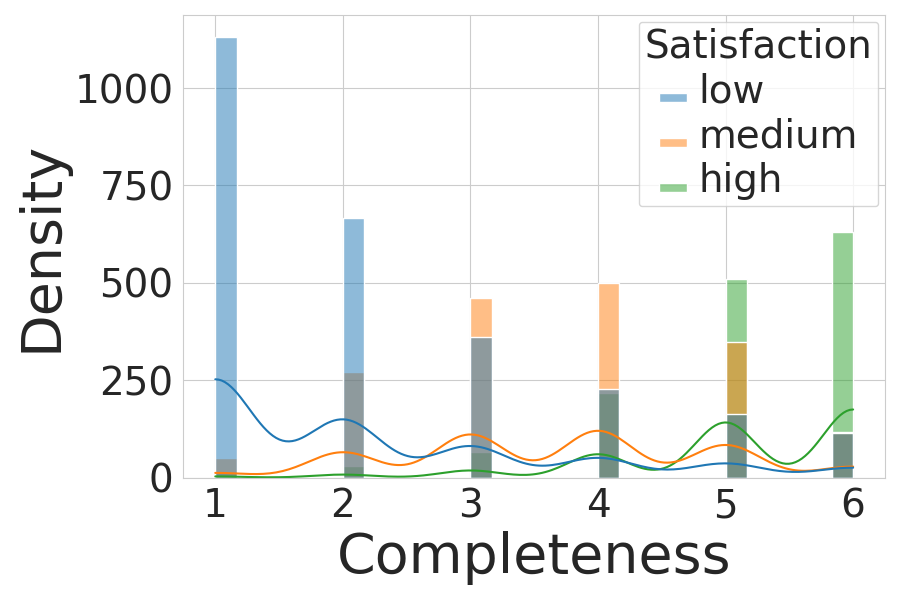}
        \caption*{}
        \label{fig:completeness_dist}
    \end{subfigure}
    \hfill
    \begin{subfigure}[t]{0.43\textwidth}
        \centering
        \includegraphics[width=\textwidth]{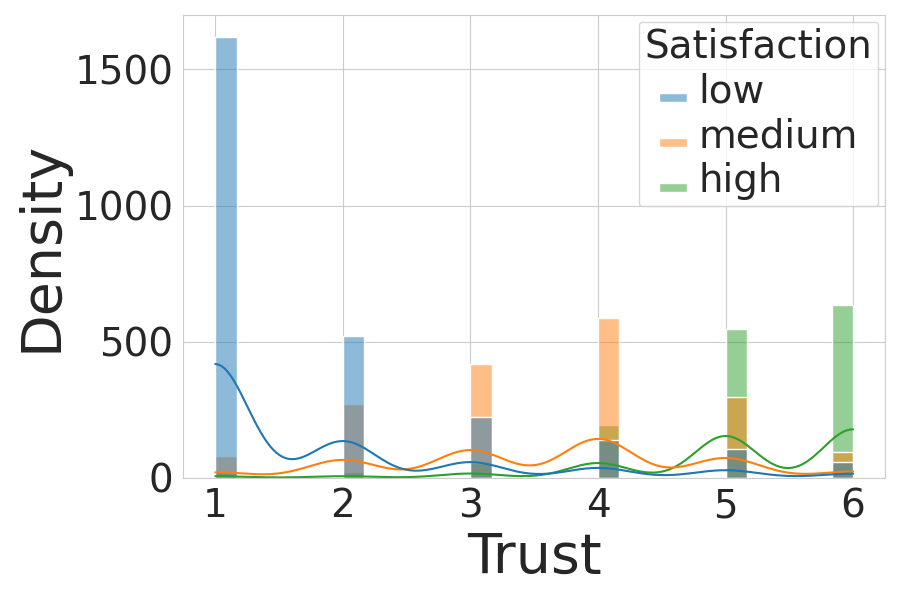}
        \caption*{}
        \label{fig:trust_dist}
    \end{subfigure}

    \begin{subfigure}[t]{0.43\textwidth}
        \centering
        \includegraphics[width=\textwidth]{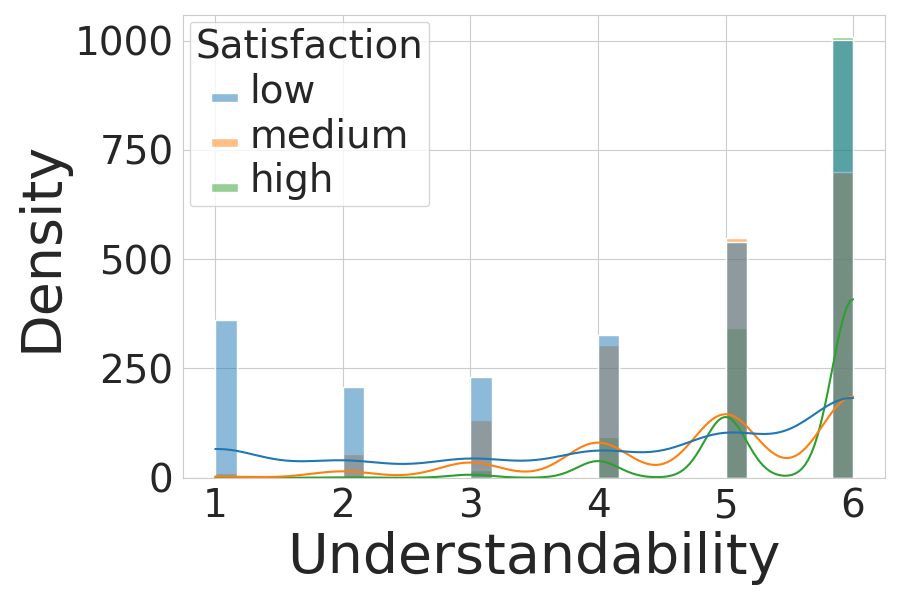}
        \caption*{}
        \label{fig:trust_dist}
    \end{subfigure}
    \hfill
    \begin{subfigure}[t]{0.43\textwidth}
        \centering
        \includegraphics[width=\textwidth]{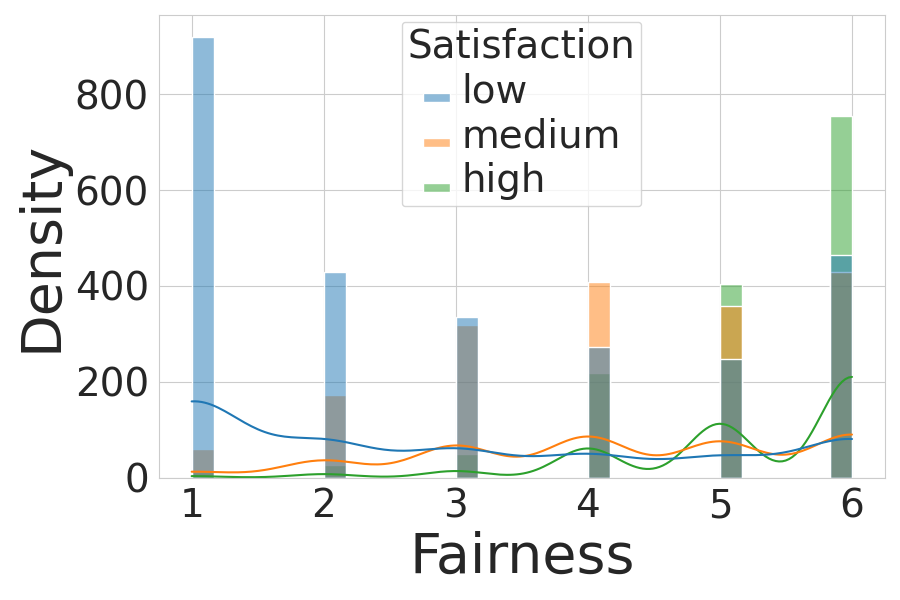}
        \caption*{}
        \label{fig:fairness_dist}
    \end{subfigure}

    \begin{subfigure}[t]{0.43\textwidth}
        \centering
        \includegraphics[width=\textwidth]{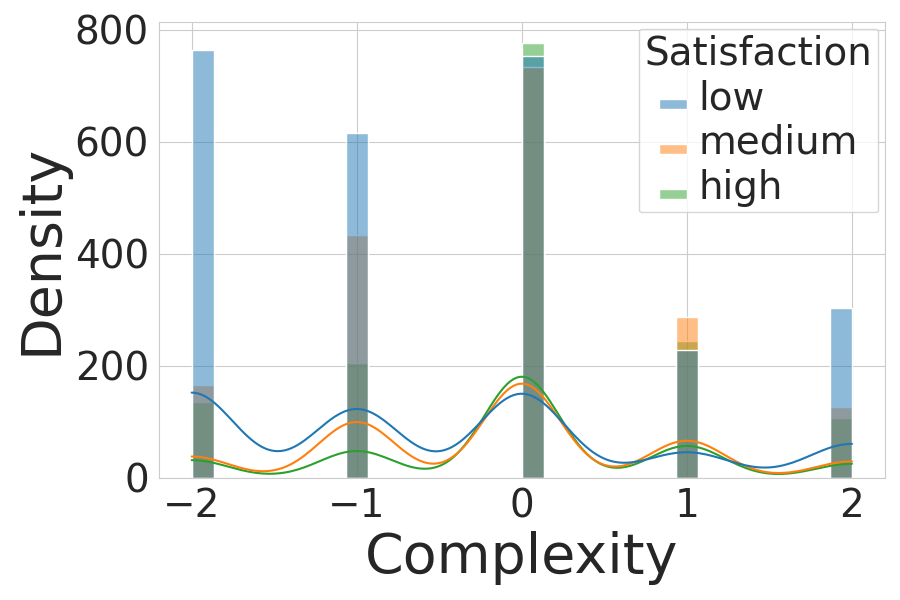}
        \caption*{}
        \label{fig:understandability_dist}
    \end{subfigure}
    \hfill
    \begin{subfigure}[t]{0.43\textwidth}
        \centering
        \includegraphics[width=\textwidth]{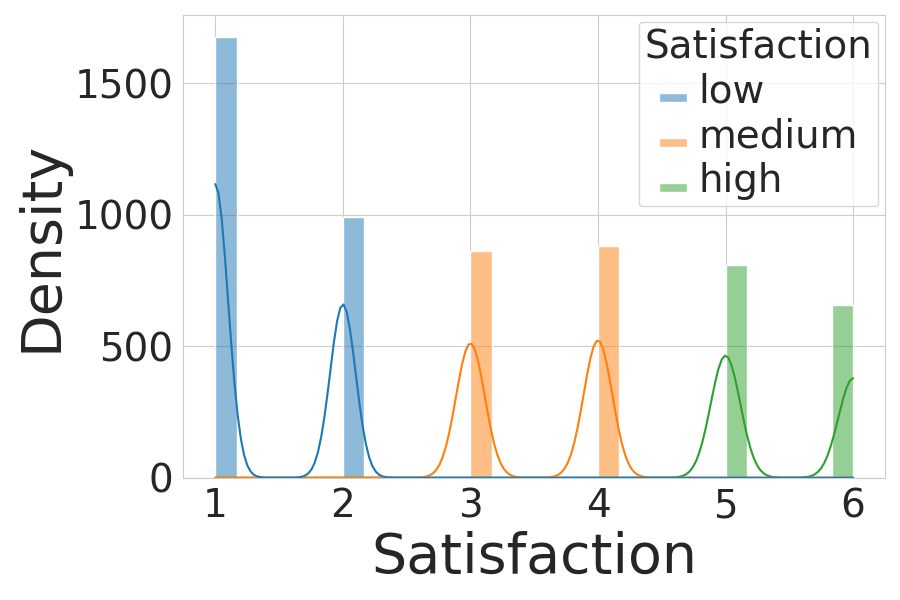}
        \caption*{}
        \label{fig:consistency_dist}
    \end{subfigure}
   
    \caption{\textbf{Per-metric distributions grouped by Satisfaction level (low, medium, high).} Each histogram is color-coded by the participant's Satisfaction category, illustrating how Feasibility, Consistency, Fairness, Completeness, Trust, Understandability, and Complexity vary for each class. 
        The final subplot depicts the overall Satisfaction distribution itself.}
    \label{fig:distribution_sat}
\end{figure}

Figure~\ref{fig:distribution_sat} shows how each metric distribution corresponds to three Overall Satisfaction categories (Low: 1–2, Medium: 3–4, High: 5–6). \textit{Feasibility}, \textit{Consistency}, \textit{Completeness}, and \textit{Trust} (all on a 1–6 scale) exhibit clear alignment with satisfaction categories: low satisfaction aligns with ratings of 1–2, high satisfaction with 5–6, and medium levels in between. \textit{Fairness} follows a similar pattern, though it exhibits a few cases of disagreement. For Complexity ($-2$ to $+2$, with $0 = desired$), high-satisfaction scenarios cluster near 0, whereas low-satisfaction scenarios are more frequent at the extremes ($-2$ = "too simple" or $+2$ = "too complex"). Understandability was defined as “I feel like I understood the phrasing of the explanation well,” and was primarily meant to filter out poor English comprehension. While most participants rated Understandability relatively high, there is still some spread possibly reflecting different interpretations of "understandable."

\subsection{Bi-clustering Analysis}

\begin{figure}[!htbp]
    \centering
    
    \begin{subfigure}[t]{0.43\textwidth}
        \centering
        \includegraphics[width=\textwidth]{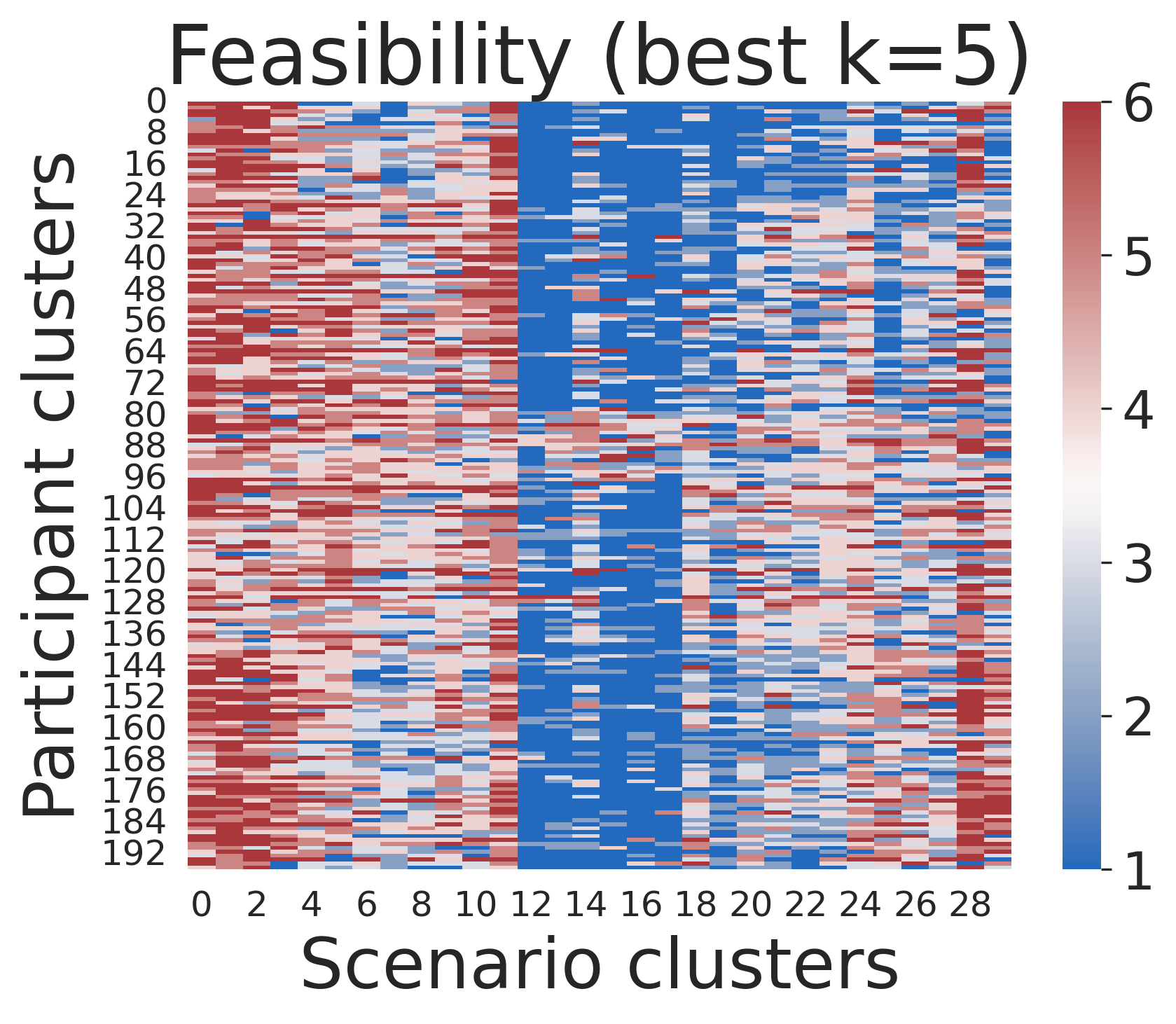}
        \caption*{}
        \label{fig:biclust_feasibility}
    \end{subfigure}
    \hfill
    \begin{subfigure}[t]{0.43\textwidth}
        \centering
        \includegraphics[width=\textwidth]{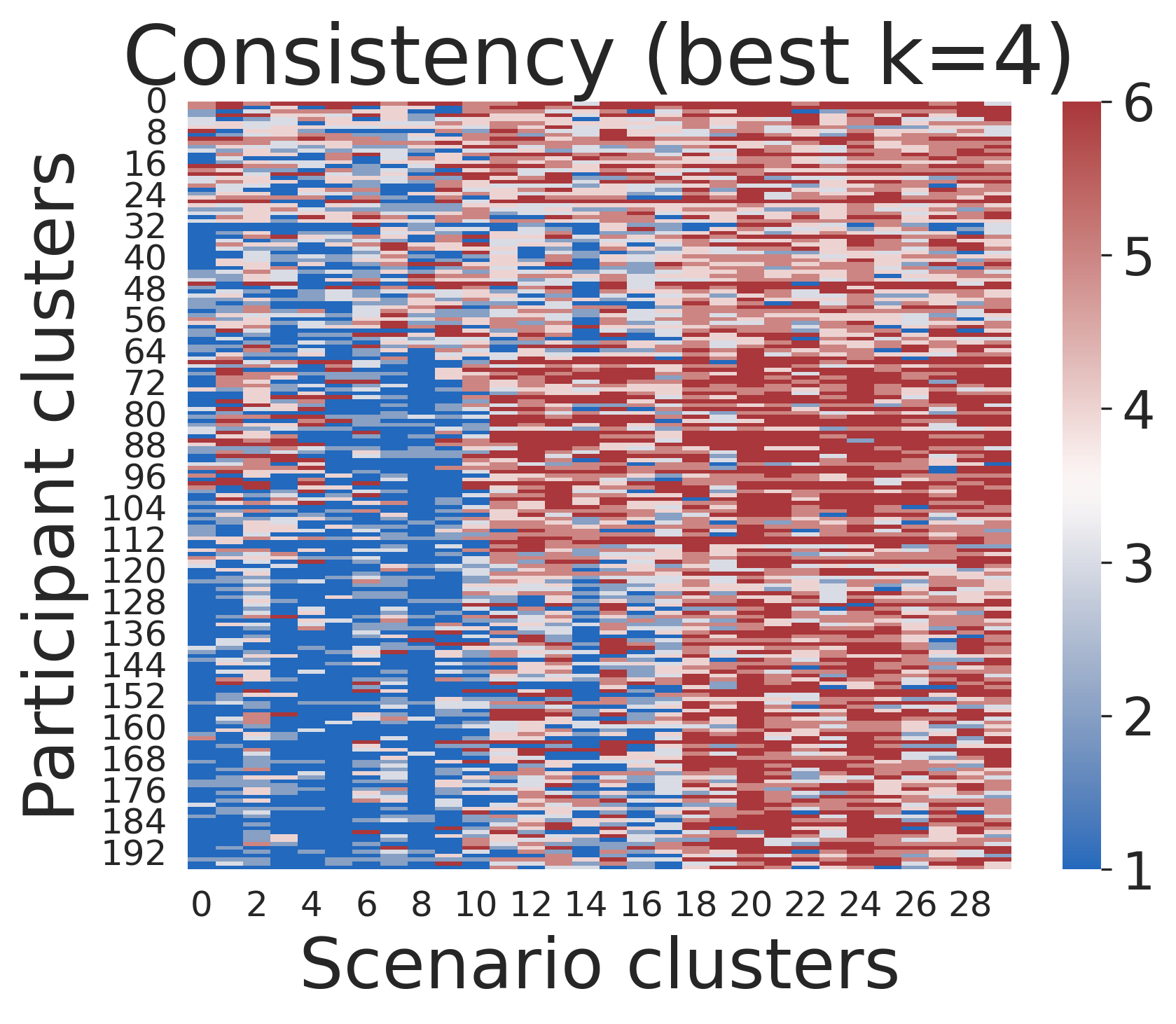}
        \caption*{}
        \label{fig:biclust_consistency}
    \end{subfigure}
    
    \vspace{-1em} 
    \begin{subfigure}[t]{0.43\textwidth}
        \centering
        \includegraphics[width=\textwidth]{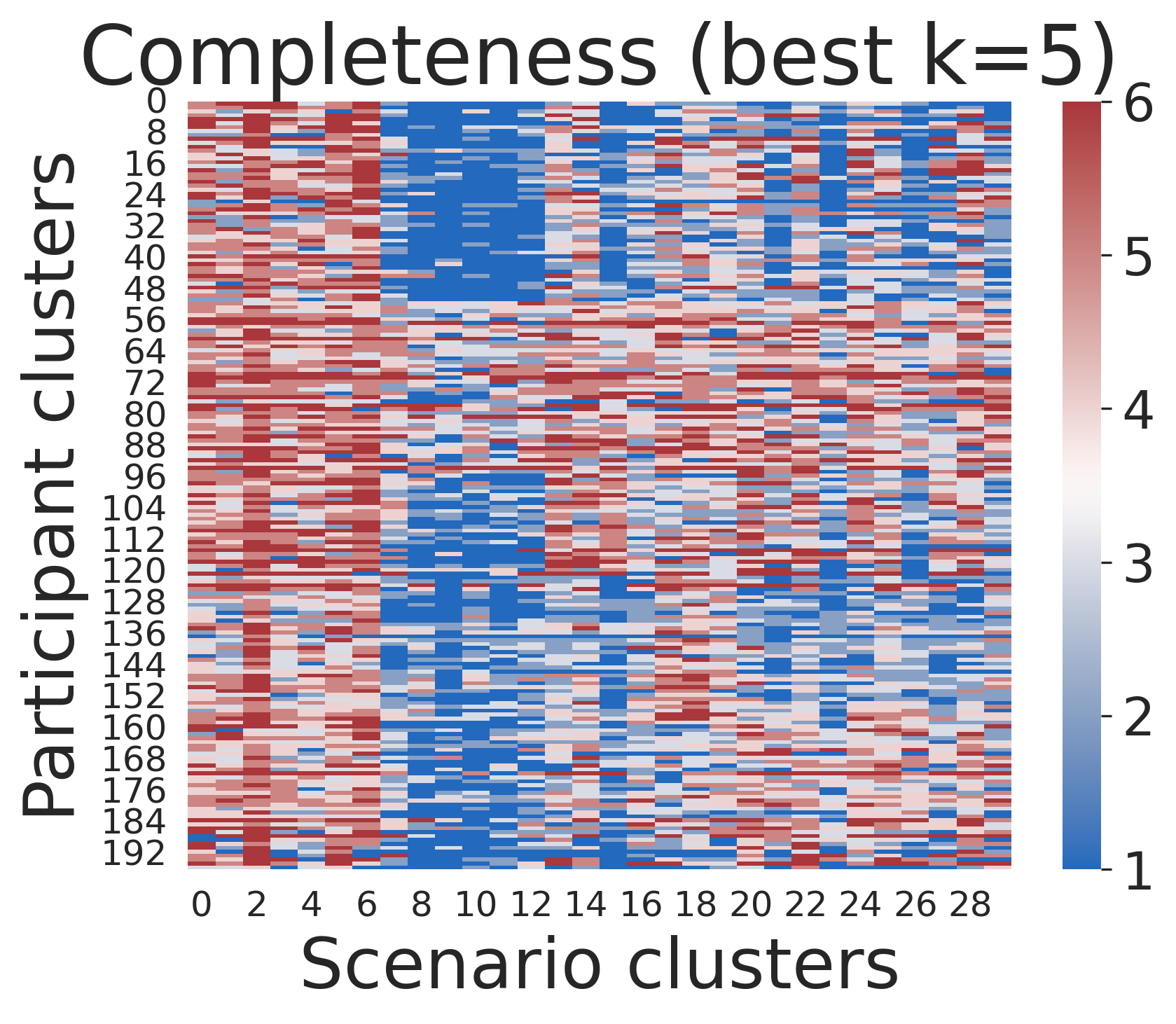}
        \caption*{}
        \label{fig:biclust_completeness}
    \end{subfigure}
    \hfill
    \begin{subfigure}[t]{0.43\textwidth}
        \centering
        \includegraphics[width=\textwidth]{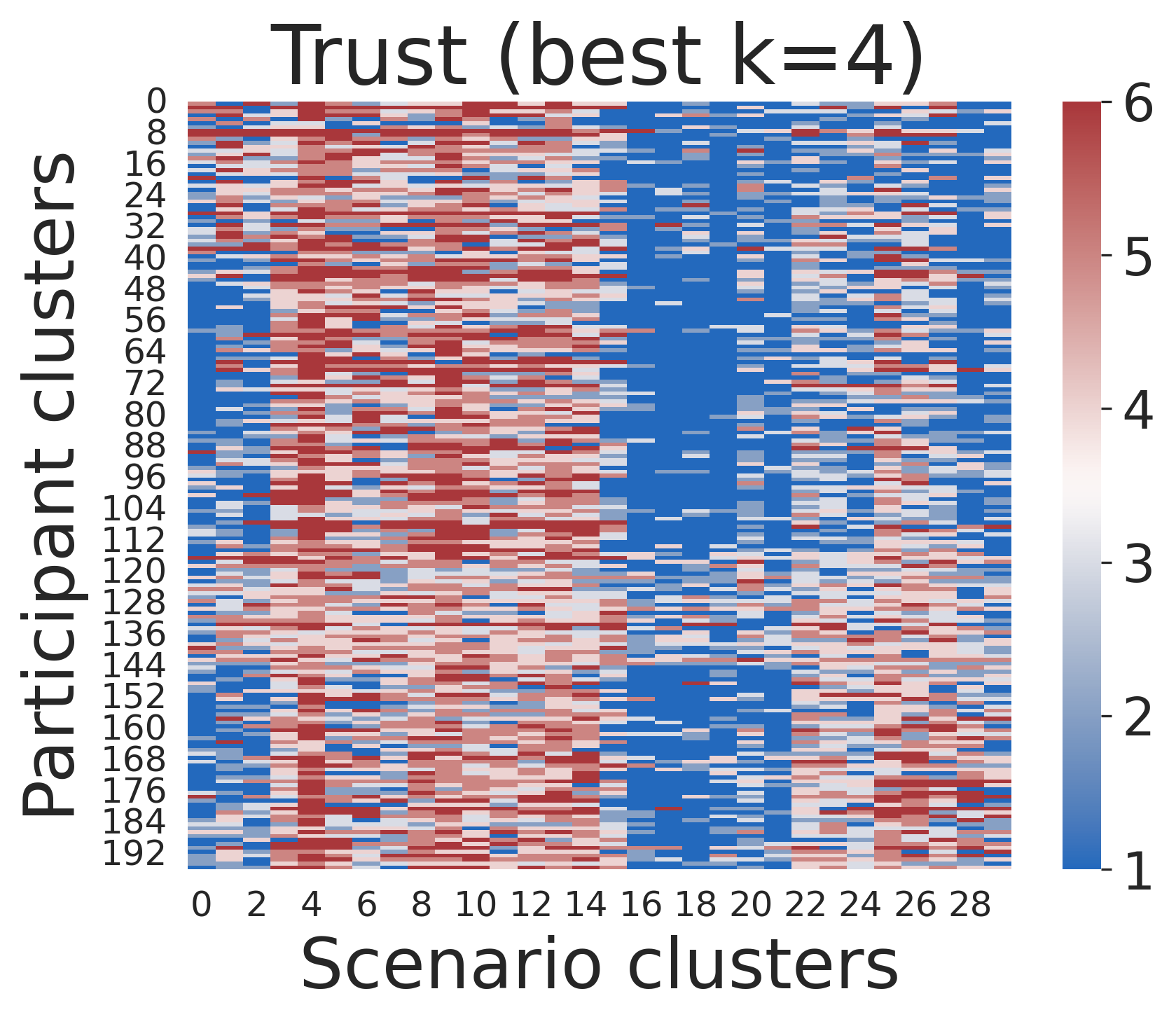}
        \caption*{}
        \label{fig:biclust_trust}
    \end{subfigure}

    \vspace{-1em}  
    \begin{subfigure}[t]{0.43\textwidth}
        \centering
        \includegraphics[width=\textwidth]{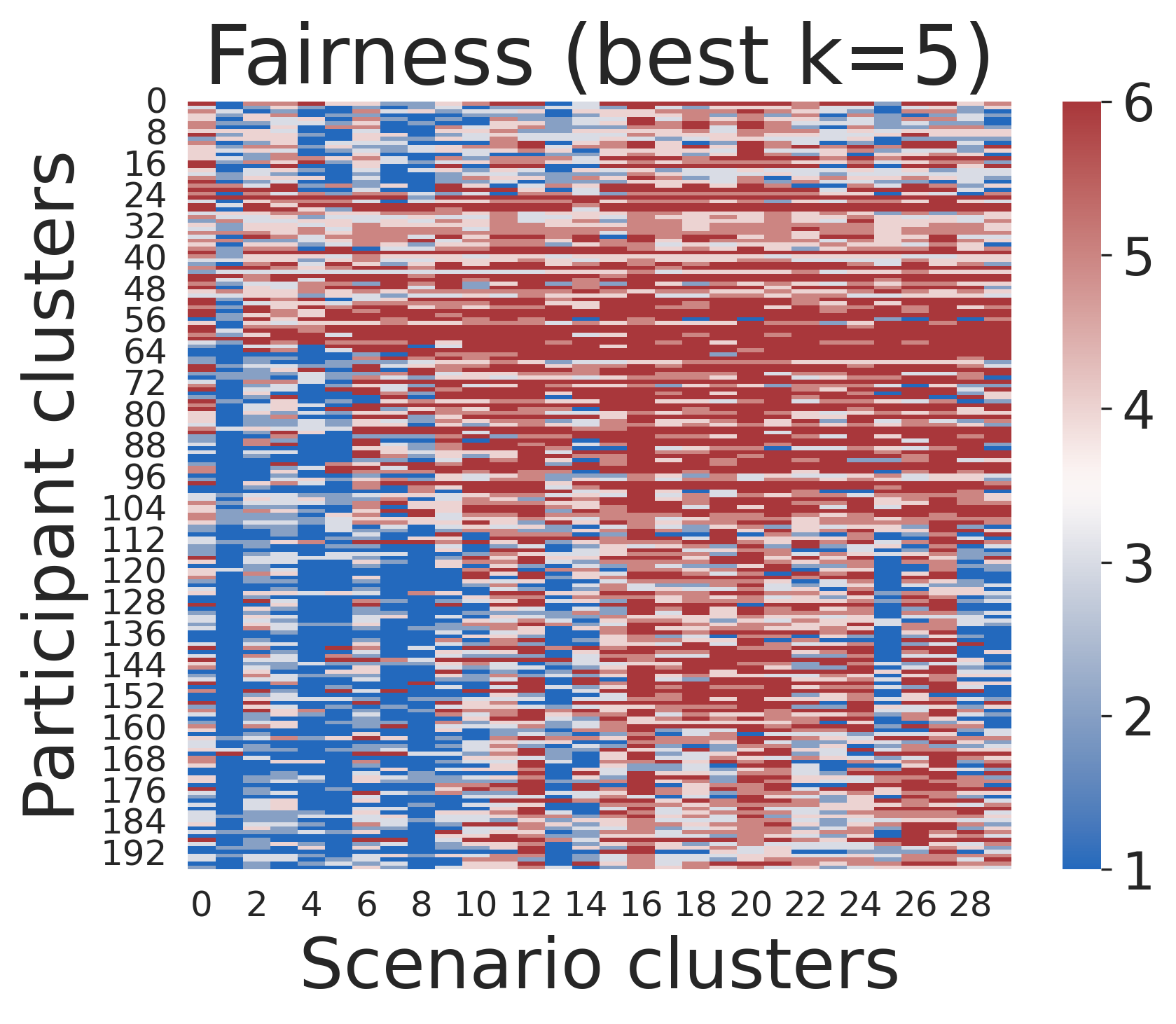}
        \caption*{}
        \label{fig:biclust_fairness}
    \end{subfigure}
    \hfill
    \begin{subfigure}[t]{0.43\textwidth}
        \centering
        \includegraphics[width=\textwidth]{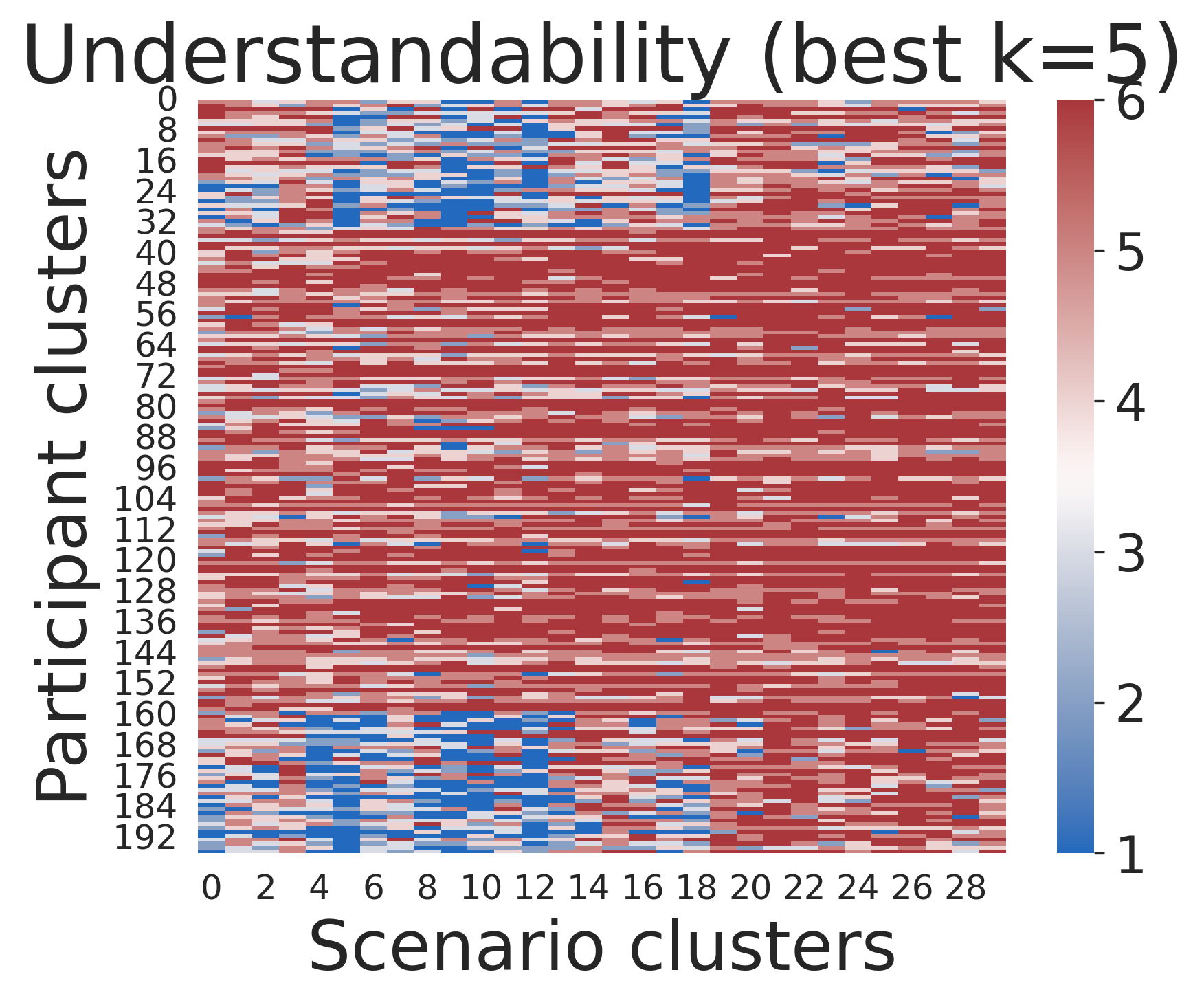}
        \caption*{}
        \label{fig:bicl_understand}
    \end{subfigure}

    \vspace{-1em}
    \begin{subfigure}[t]{0.43\textwidth}
        \centering
        \includegraphics[width=\textwidth]{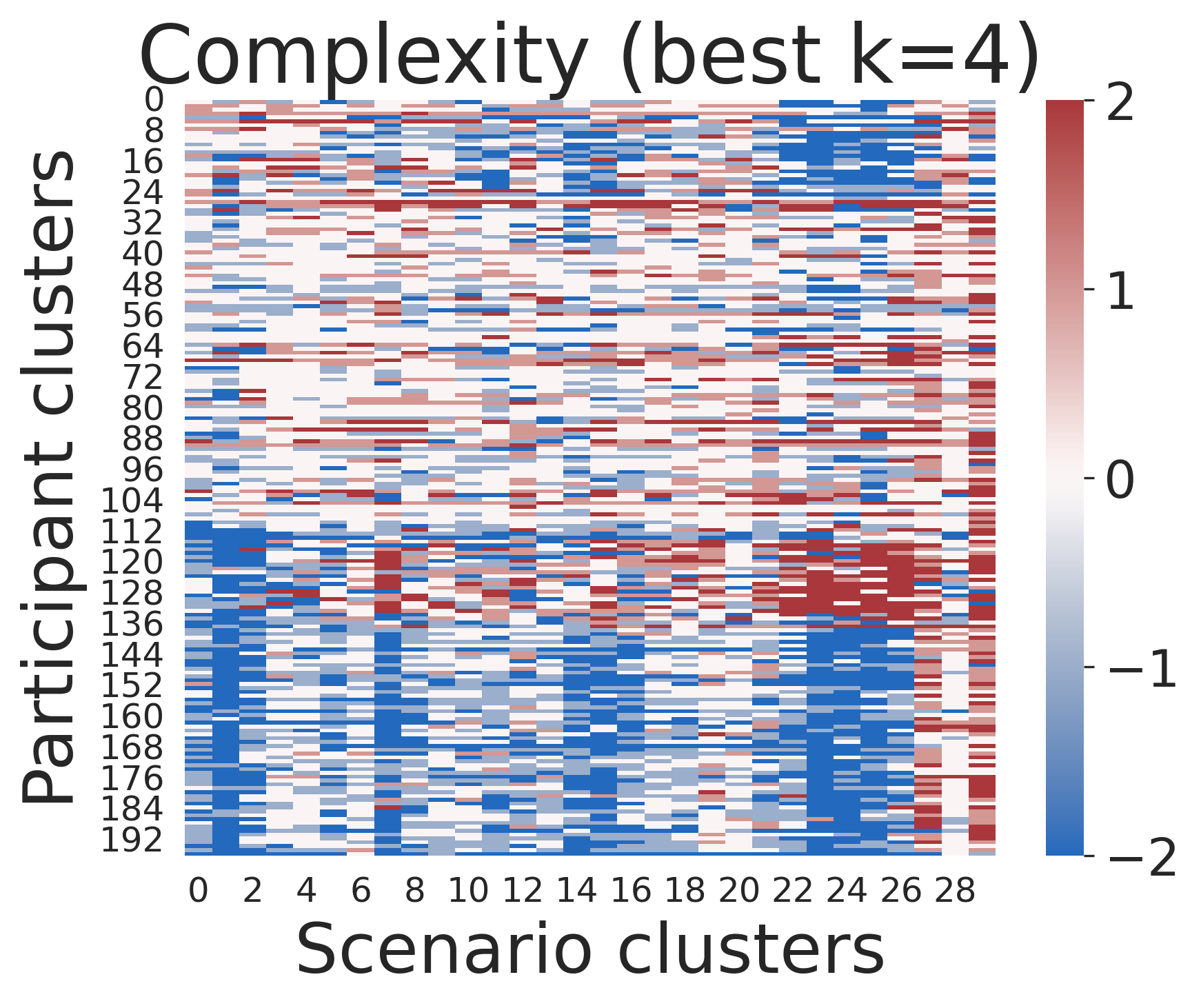}
        \caption*{}
        \label{fig:understandability_dist}
    \end{subfigure}
    \hfill
    \begin{subfigure}[t]{0.43\textwidth}
        \centering
        \includegraphics[width=\textwidth]{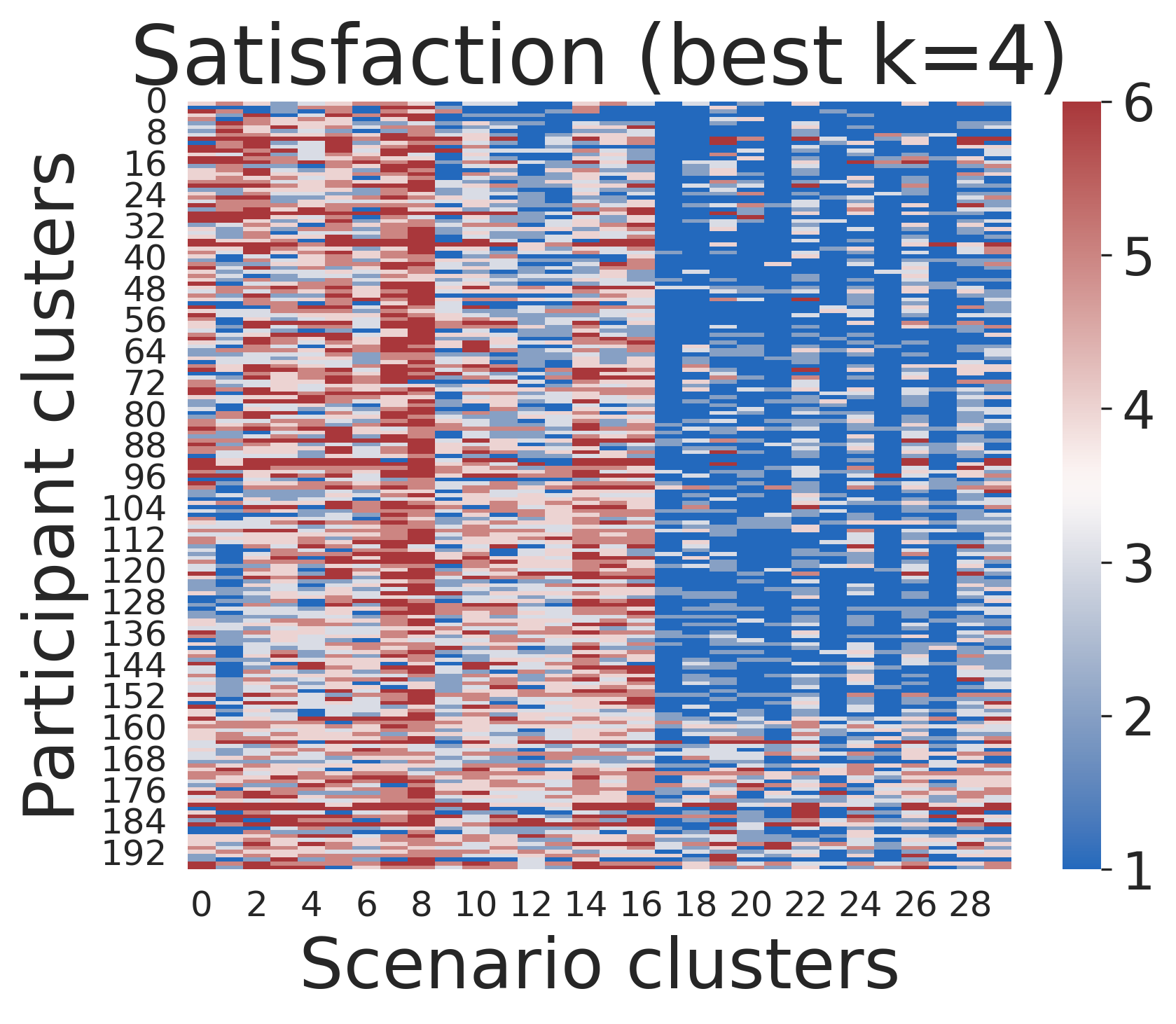}
        \caption*{}
        \label{fig:bicl_consistency}
    \end{subfigure}
   
    \caption{\textbf{Bi-clustering results for all metrics.} 
Each heatmap shows participants (rows) and scenarios (columns) reordered into \(k\) co-clusters, where \(k\) is selected by minimizing the reconstruction error.}
    \label{fig:biclustering}
\end{figure}

To further explore latent patterns of agreement among participants and scenarios, we performed a bi-clustering (simultaneously clustering of rows = participants, and columns = scenarios) using each metric’s rating matrix (Figure~\ref{fig:biclustering}). We applied Spectral Co-clustering to identify clear and distinct groups of participants (who systematically differ in their evaluation patterns) and explanatory metrics (rated similarly by certain groups of participants). We varied the number of clusters from 2 to 6, choosing the best $k$ to minimize the Frobenius norm (reconstruction error). Several metrics, such as Feasibility, Completeness, and Fairness, favor a 5-cluster, while others (e.g., Consistency, Trust, Complexity, Satisfaction) align better with 4. 

Vertically, distinct scenario clusters emerge, indicating groups of scenarios that receive systematically higher or lower scores for a given metric. In the \textit{Satisfaction} subplot, for instance, some columns (scenarios) are predominantly blue, indicating many participants deem them highly satisfactory, while others lean more red, reflecting lower satisfaction across most evaluators. \textit{Feasibility} and \textit{Trust} similarly show strong column‐based patterns, suggesting consensus among participants on which scenarios appear feasible or trustworthy.

Horizontally, we see participant clusters signifying "generally high raters," "generally low raters," or more moderate profiles. \textit{Complexity}, in particular, exhibits horizontal stripes, implying that participants differ in perceiving an explanation’s complexity, some appear to rate nearly all scenarios as highly complex, while others rate them as simpler. We further refine this intuition by checking evaluators' background data.

For \textit{Consistency} and \textit{Completeness}, the presence of red columns suggests that when a scenario is perceived as incomplete or inconsistent, most participants converge on low scores, while for other scenarios some graders remain more conservative, with only a few having clear vertical blue line. A similar dynamic arises for \textit{Fairness}: highly biased scenarios provoke universal low ratings, while more ambiguous cases show greater disagreement. Meanwhile, \textit{Understandability} ratings are high, indicating that participants typically grasp the definition of "understandable," yet a few row deviate, suggesting different interpretation.

It is worth noting that all scenarios were presented in random order, so these vertical clusters do not reflect any intended sequence or grouping in the survey.

\subsection{Demographic Influences}
\begin{figure}
    \centering
    
    \begin{subfigure}[t]{0.45\textwidth}
        \centering
        \includegraphics[width=\textwidth]{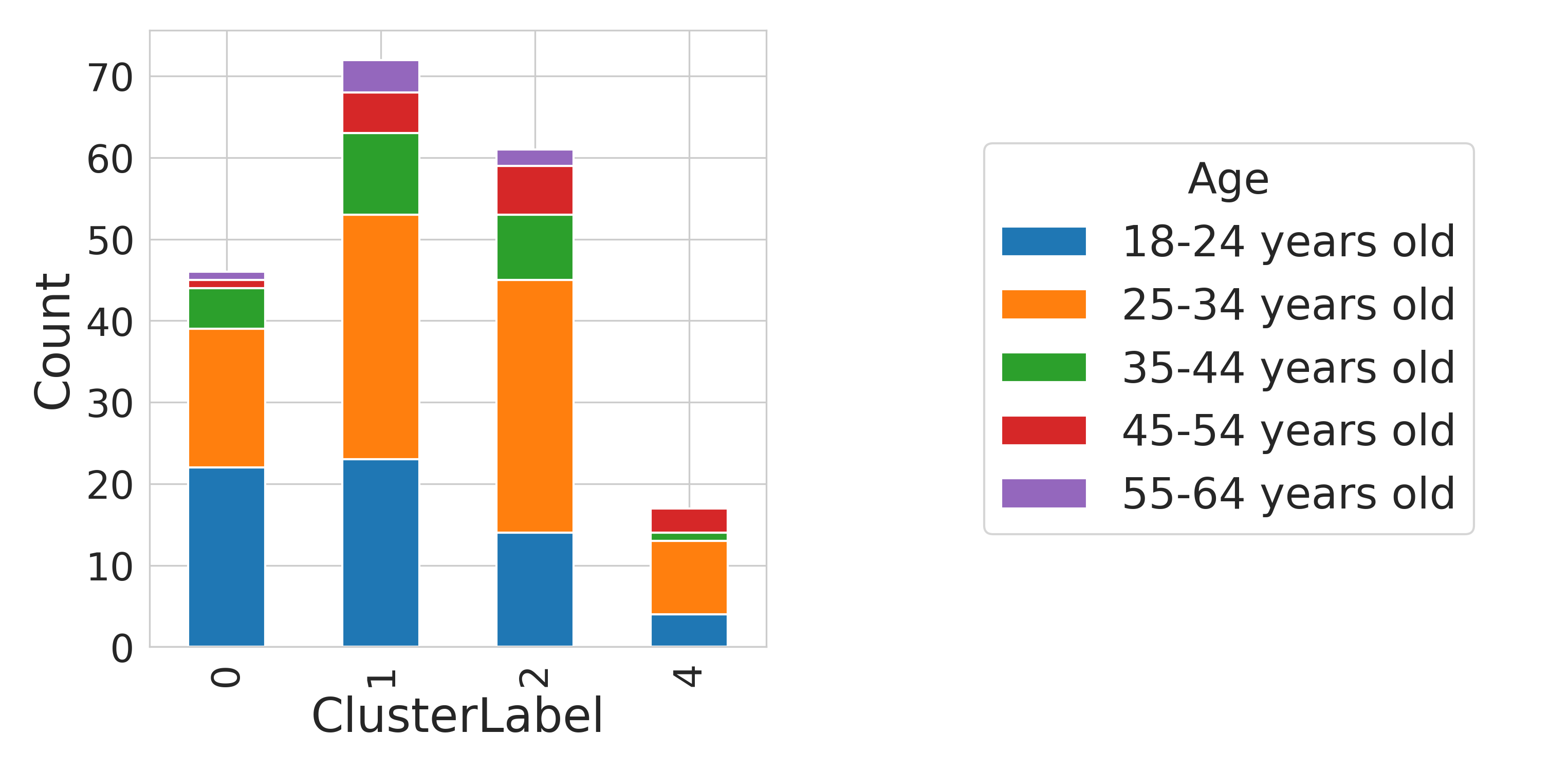}
        \caption{\textbf{Age vs.\ ClusterLabel.} 
        \textit{Chi-square=14.005, p=0.3004, dof=12. 
        Age distribution does not differ significantly across clusters.}}
    \end{subfigure}
    \hfill
    \begin{subfigure}[t]{0.45\textwidth}
        \centering
        \includegraphics[width=\textwidth]{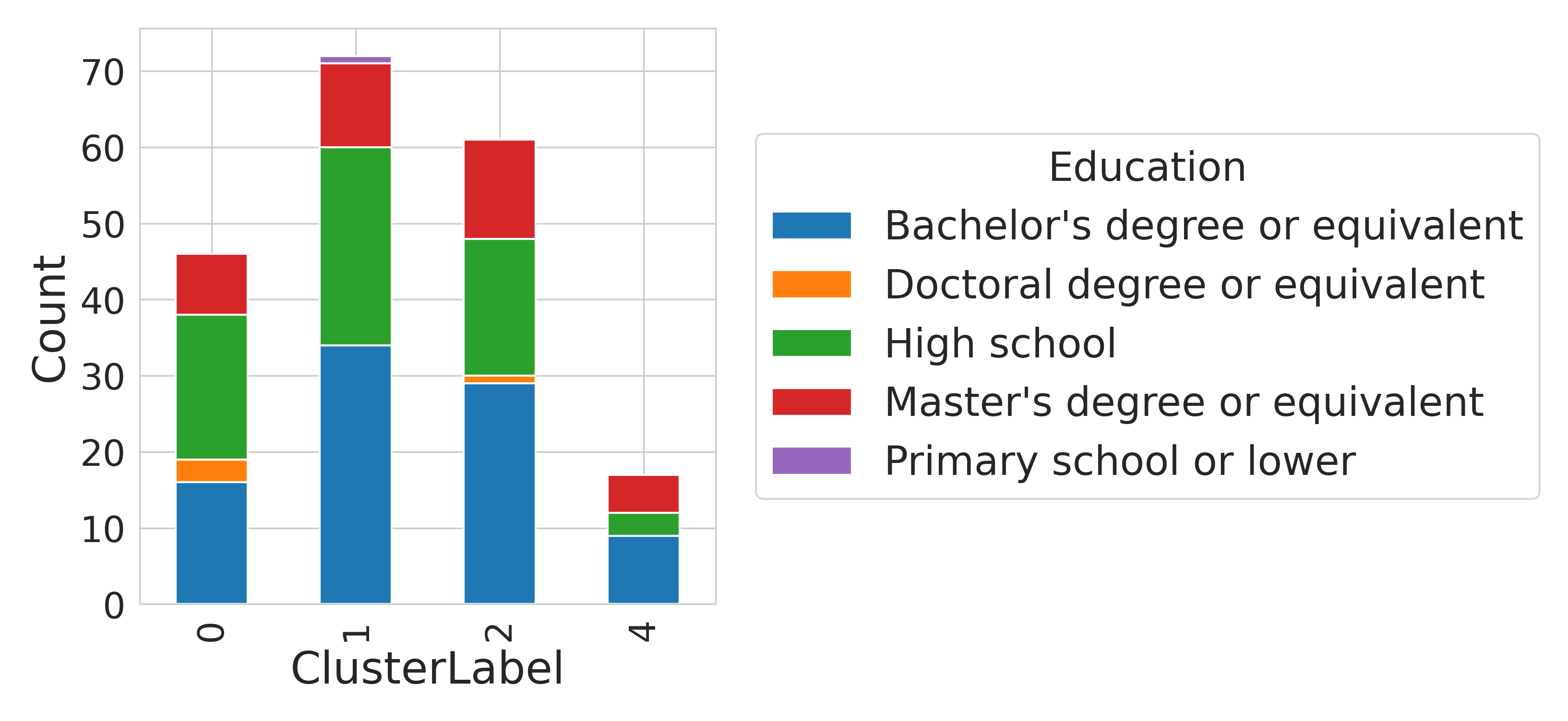}
        \caption{\textbf{Education vs.\ ClusterLabel.} 
        \textit{Chi-square=13.860, p=0.3098, dof=12. 
        No significant difference in education across clusters.}}
    \end{subfigure}
    
    \vspace{1em}
    
    \begin{subfigure}[t]{0.45\textwidth}
        \centering
        \includegraphics[width=\textwidth]{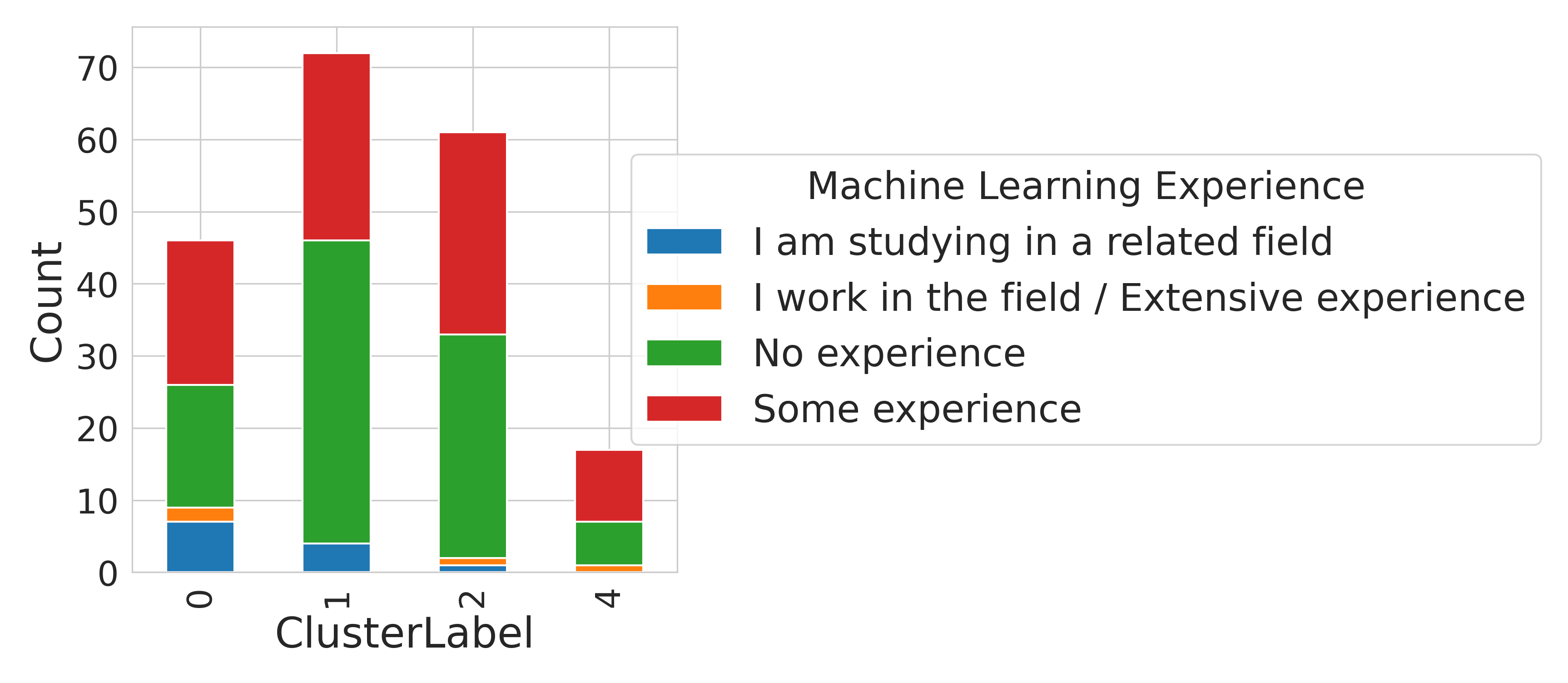}
        \caption{\textbf{MLExperience vs.\ ClusterLabel.} 
        \textit{Chi-square=18.490, p=0.0299, dof=9. 
        Significant difference in ML experience across clusters.}}
    \end{subfigure}
    \hfill
    \begin{subfigure}[t]{0.45\textwidth}
        \centering
        \includegraphics[width=\textwidth]{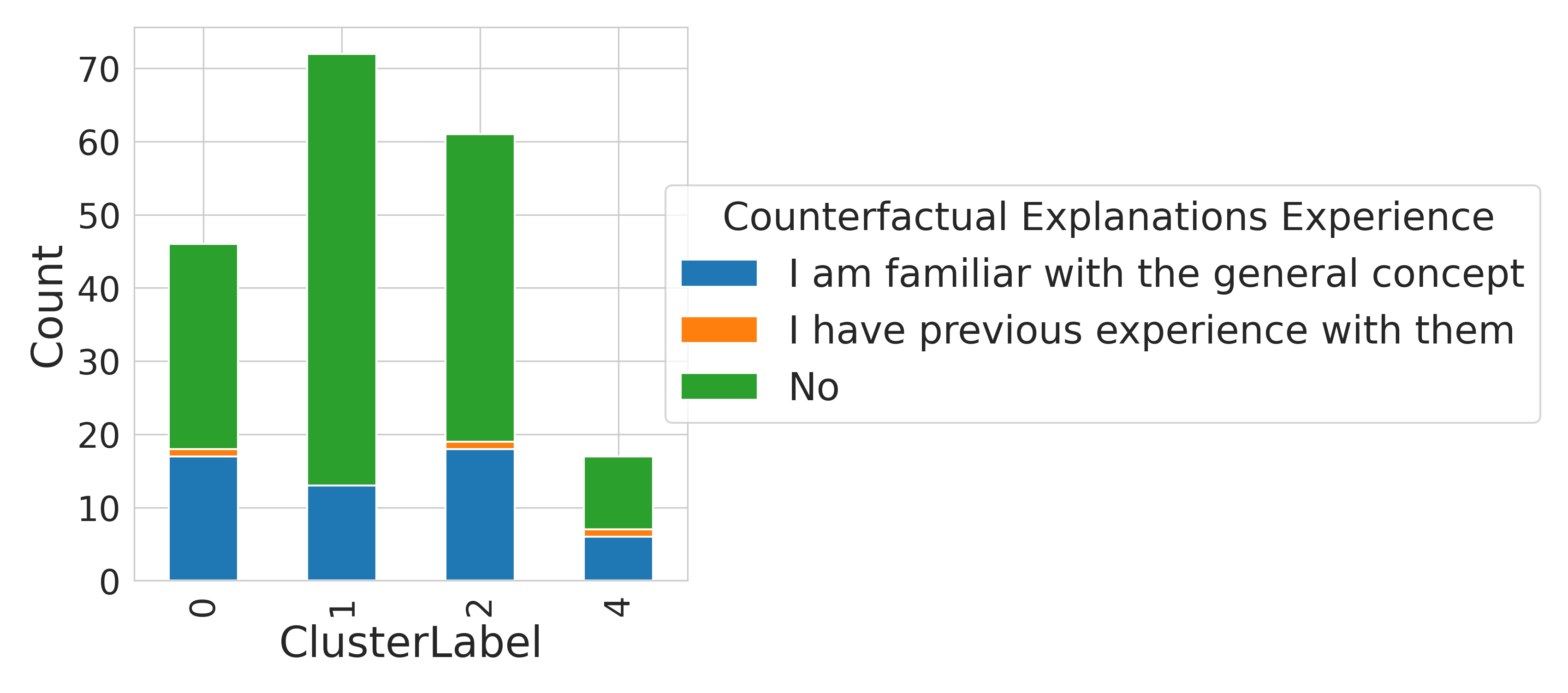}
        \caption{\textbf{CFEExperience vs.\ ClusterLabel.} 
        \textit{Chi-square=9.899, p=0.1290, dof=6. 
        Not significant for counterfactual frameworks experience.}}
    \end{subfigure}
    
    \vspace{1em}
    
    \begin{subfigure}[t]{0.45\textwidth}
        \centering
        \includegraphics[width=\textwidth]{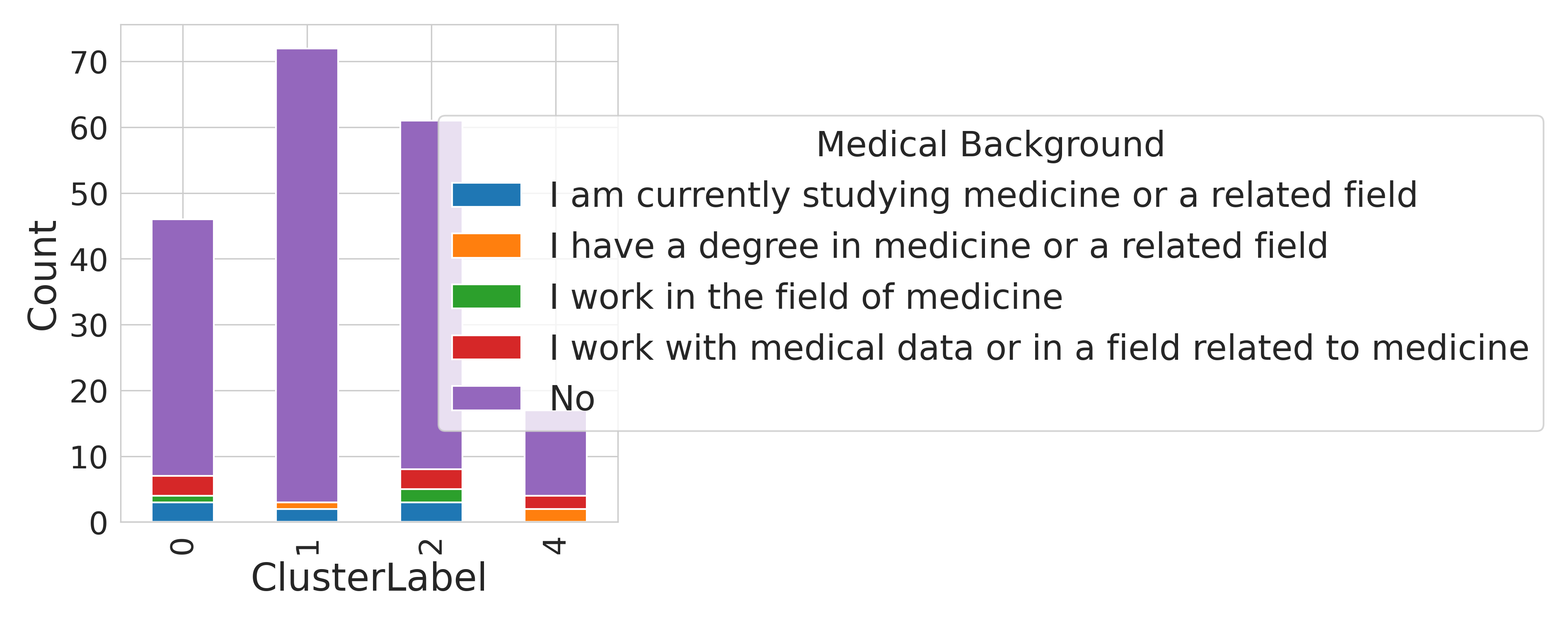}
        \caption{\textbf{MedicalBackground vs.\ ClusterLabel.} 
        \textit{Chi-square=24.728, p=0.0162, dof=12.
        Significantly different across clusters.}}
    \end{subfigure}
    \hfill
    \begin{subfigure}[t]{0.45\textwidth}
        \centering
        \includegraphics[width=\textwidth]{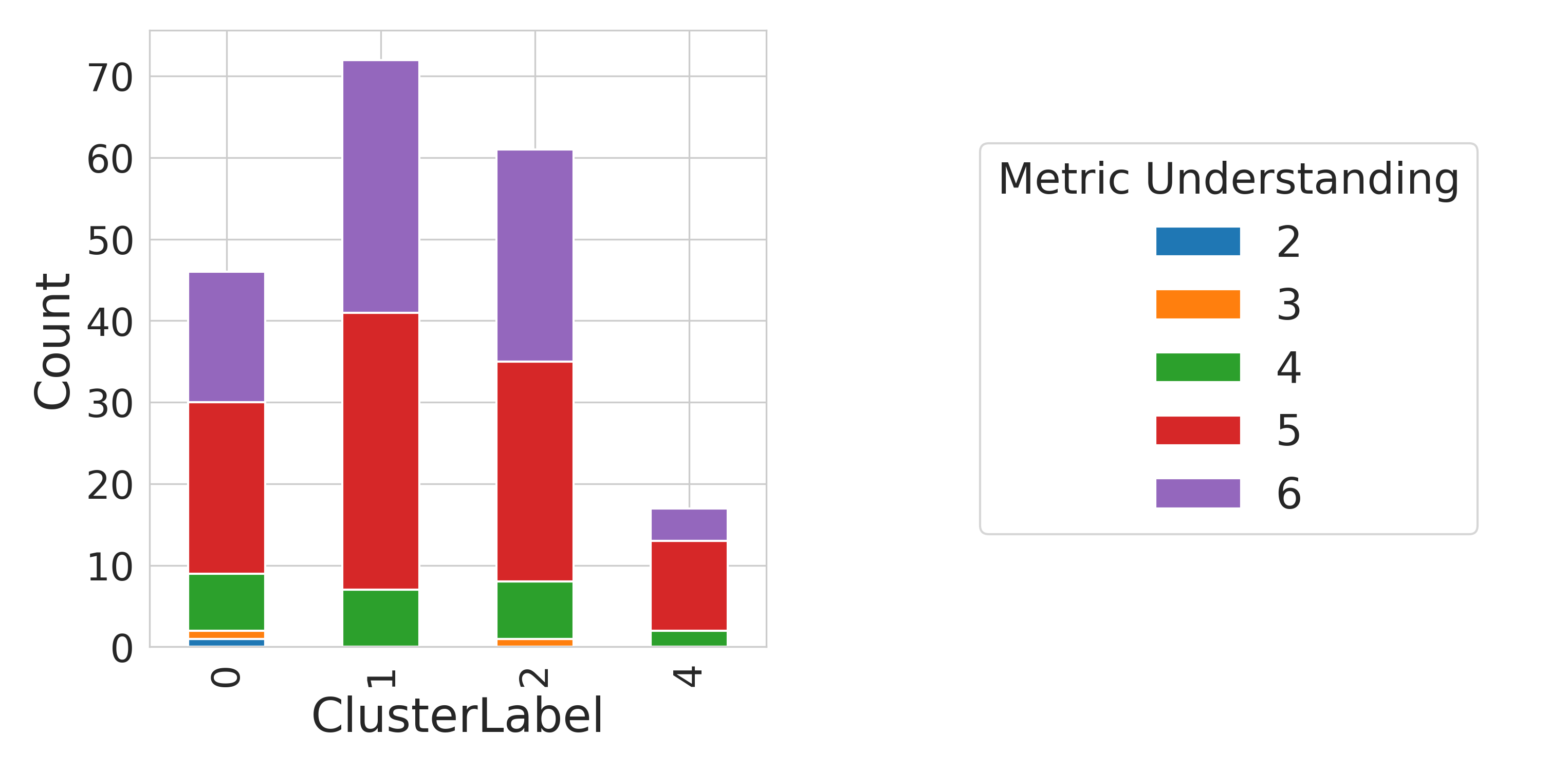}
        \caption{\textbf{MetricUnderstanding vs.\ ClusterLabel.}
        \textit{Chi-square=8.700, p=0.7283, dof=12.
        No significant difference in how well participants understood metrics.}}
    \end{subfigure}
    
    \caption{\textbf{Distribution of evaluators' background (Age, Education, Machine Learning Experience, etc.) within the four discovered clusters}. 
    Chi-square tests (with p-values and degrees of freedom) assess whether the distributions differ significantly among clusters. 
    We find that Machine Learning Experience (\textit{p=0.0299}) and Medical Background (\textit{p=0.0162}) differ significantly across clusters, whereas Age, Education, Counterfactual Explanations Experience, and Metric Understanding do not show significant differences.}
    \label{fig:demographics}
\end{figure}

To see whether respondent backgrounds explained these cluster memberships, we analyzed demographic and experience‐related data (e.g., age, education, ML experience). As shown in Figure~\ref{fig:demographics}, \textbf{Machine Learning Experience} and \textbf{Medical Background} emerged as the most relevant demographic factors, both showing statistically significant differences  across the clusters ($p=0.0299$ and $p = 0.0162$, respectively). This indicates that domain experts often require more thorough justification before trusting computational outputs \cite{bansal2019beyond}, particularly in healthcare context \cite{ghassemi2021false}. It is also evident that people with Machine Learning knowledge tend to evaluate model reasoning more critically, displaying under-reliance \cite{10.1145/3613904.3642474}. On the other hand, Age, Education did not differ significantly across bi-clusters ($p>0.05$). Likewise, Metric Understanding showed no major effect, suggesting that while participants’ perceived familiarity with the metrics varied (indicating confidence in their understanding), their overall ranking behavior remained aligned. Lastly, no sufficient evidence emerged to link Counterfactual Explanation familiarity to specific rating patterns, likely due to the small proportion of participants reporting such experience.

\subsection{Factor Analysis} 
To investigate the underlying structure among the seven metrics, we concluded an exploratory factor analysis. We used Bartlett's test and the Kaiser-Meyer-Olkin (KMO) measure of sampling adequacy to verify if factor analysis is appropriate. The Bartlett test was highly significant  ($\upchi^2=28150.28$, $p<0.001$),  indicating that the correlation matrix is not an identity matrix, while the KMO value of 0.893 indicating sampling adequacy.  A scree plot of the eigenvalues suggested three factors, as there was a clear “elbow” after the third component. Table ~\ref{factor} shows the rotated (varimax) factor loadings for the three-factor solution. The first factor alone accounts for approximately  40.5\% of the total variance, the second factor brings the cumulative variance to  56.9\%, and the third factor increases it to 
65.9\%. Evidently, several metrics (e.g., Feasibility, Consistency, Trust) load strongly on the first factor, whereas Understandability loads most heavily on the second factor. The third factor has moderate loadings (up to 0.5719), indicating a separate, though smaller, source of variance among the items.

\begin{table}[H]
\caption{Varimax-rotated loadings of the first three extracted factors from a factor analysis on the seven metrics (Feasibility, Consistency, Completeness, Trust, Understandability, Fairness, and Complexity). The final column shows the cumulative percentage of variance explained by the corresponding factor.}\label{factor}
\centering
\begin{tabular}{|l|ccccccc|l|}
\hline
\textbf{Factor} &  Feasib. & Consist. & Complet. & Trust & Understand. & Fairness & Complex. & Cumul. var.\\
\hline
0 &  0.7805 & 0.7697 & 0.6273 & 0.7896 & 0.3721 & 0.6884 & 0.0319 & 40.5\%\\
1 &  0.1543 & 0.2696 & 0.2820 & 0.1846 & 0.9200 & 0.2965 & -0.0612 & 56.9\%\\
2 & 0.1018 & 0.1628 & 0.5719 & 0.3332 & -0.1031 & -0.0688 & 0.3778 & 65.9\%\\
\hline
\end{tabular}
\end{table}

\section{Predicting Overall Satisfaction with Machine Learning Methods}

We employed a series of supervised machine learning methods to model overall satisfaction. Satisfaction was modeled as both regression (treating it as continuous outcome) and classification (categorizing it into low, medium, high). The dataset cube consists of 30 scenarios and 196 evaluators, each providing ratings on seven explanatory metrics plus an overall satisfaction score, resulting in 5,880 total instances. Notably, demographic variables were not used as predictive features. We then applied two data-splitting strategies:

\begin{enumerate}
    \item \textit{Random Split}: We randomly partitioned the 5,880 instances into 80\% training set and a 20\% test set, then performed 5-fold cross-validation on the training subset.   
    \item \textit{Scenario-Based Split}: We assigned entire scenarios to either training or test sets, with 24 scenarios (24\(\times\)196 = 4,704 instances) for training and 6 scenarios (6\(\times\)196 = 1,176 instances) for testing. This ensures the model is evaluated on completely unseen scenarios.
\end{enumerate}

\subsection{Predicting Overall Satisfaction as Regression Problem}
\subsubsection{Estimating Coefficients with OLS Regression}

We then used ordinary least squares (OLS) regression to model Overall Satisfaction directly. 
Table ~\ref{tab:ols_regression} presents the estimated coefficients, standard errors, and confidence intervals. The intercept is small but significant ($0.1766$), forming a baseline for Satisfaction when all metrics are zero. Overall, the OLS model achieves a high \textbf{$R^2= 0.757 \pm 0.008$}, meaning these metrics collectively explain a substantial portion of participants’ Satisfaction ratings variation. \emph{Feasibility} ($\beta \approx 0.358$) and \emph{Trust} ($\beta \approx 0.362$) bring the largest positive effects: a 1-point increase in either metric (while holding all other metrics constant) corresponds to approximately a 0.36-point increase in Satisfaction. \emph{Completeness} also shows a substantial positive coefficient ($\beta = 0.170; p<0.001$), while \emph{Consistency} ($\beta = 0.067$) and \emph{Complexity} ($\beta = 0.080$) have somewhat smaller (but still significant) positive effects.  

\begin{table}[H]
\centering
\caption{\textbf{OLS regression results modeling Overall Satisfaction.}
Reported are the coefficient estimates, standard errors, t-values, p-values, and 95\% confidence intervals for each predictor.}
\begin{tabular}{|l|c|cccc|}
\hline
\bfseries Predictor         &  \bfseries Coefficient & \bfseries Std. Error & \bfseries t-value & \bfseries p-value & \bfseries 95\% CI \\
\hline
Intercept         & 0.1766      & 0.040      & 4.410   & 0.000   & [0.098, 0.255] \\
Feasibility       & \textbf{0.3581}      & 0.010      & 36.563  & 0.000   & [0.339, 0.377] \\
Consistency       & 0.0665      & 0.010      & 6.347   & 0.000   & [0.046, 0.087] \\
Completeness      & 0.1702      & 0.011      & 16.144  & 0.000   & [0.150, 0.191] \\
Trust             & \textbf{0.3618}      & 0.011      & 31.796  & 0.000   & [0.340, 0.384] \\
Understandability & $-$0.0690  & 0.010      & $-$7.036& 0.000   & [$-$0.088, $-$0.050] \\
Fairness          & 0.0170      & 0.009      & 1.908   & 0.056   & [$-$0.000, 0.034] \\
Complexity        & 0.0802      & 0.010      & 7.658   & 0.000   & [0.060, 0.101] \\
\hline
\end{tabular}
\label{tab:ols_regression}
\end{table}

By contrast, \emph{Understandability} ($\beta \approx -0.069$) exhibits a negative sign, suggesting that when other metrics are accounted for, higher Understandability ratings slightly reduce predicted Satisfaction. Given the correlated nature of these metrics, the coefficient reflects the complex overlaps between Understandability and the other factors. Finally, \emph{Fairness} is borderline significant ($p\approx0.056$, $\beta=0.017$), with a 95\% CI that narrowly includes zero, suggesting it adds limited explanatory power once Feasibility, Trust, and the other predictors are included.

\subsubsection{Omitting Feasibility and Trust}
To assess the relative importance of \emph{Feasibility} and \emph{Trust}, we reran the training after removing them from the predictor set. This reduced model’s performance dropped to $R^2 = 0.580 \pm 0.081$, underscoring Feasibility and Trust’s critical role in explaining Satisfaction. In the reduced specification, \emph{Completeness} ($\beta \approx 0.413$) becomes the strongest driver, followed by \emph{Consistency} ($\beta \approx 0.322$), \emph{Fairness} ($\beta \approx 0.182$), \emph{Complexity} ($\beta \approx 0.096$), and \emph{Understandability} remains negative ($\beta \approx -0.082$) with all coefficients being significant. Nonetheless, even without Feasibility and Trust the model still captures around 58\% of Satisfaction’s variance, meaning that other factors (especially Completeness) also play a notable part. In practical terms, while enhancing Feasibility and building Trust in such explanations are the most important for maximizing user satisfaction, addressing Completeness, Consistency, and Complexity can further refine users’ overall satisfaction.

\subsubsection{Comparing Regression Models}
We next compared Linear Regression (OLS), Decision Tree Regressor, and Random Forest Regressor under both data-splitting strategies using 5-fold cross-validation. Table~\ref{tab:model_comparison_2splits} shows that \textit{Linear Regression} achieved 
RMSE $\approx 0.890 \pm 0.017$ (random split) and $0.891 \pm 0.071$ (scenario-based), with $R^2 \approx 0.73$. Decision Tree slightly improves upon OLS in the random split 
($\text{RMSE} \approx 0.859 \pm 0.030, R^2 \approx 0.755 \pm 0.016$) 
but performs worse in the scenario-based split 
($\text{RMSE} \approx 1.128, R^2 \approx 0.549$). 
This indicates that the Decision Tree may overfit specific scenarios. \textit{Random Forest} achieves similar results to Linear Regression on both splits.
  
\begin{table}[h!]
\centering
\caption{Comparison of regression model performances in predicting satisfaction using 5-fold cross-validation for two data-splitting strategies: random split and scenario-based split. Reported metrics are RMSE and $R^2$, presented as mean $\pm$ standard deviation.}
\label{tab:model_comparison_2splits}
\begin{tabular}{|l|cc|cc|}
\hline
\multirow{2}{*}{\textbf{Model}} 
  & \multicolumn{2}{c|}{\textbf{Random Split}} 
  & \multicolumn{2}{c|}{\textbf{Scenario-Based Split}} \\

& \textbf{RMSE} 
& \textbf{R\textsuperscript{2}} 
& \textbf{RMSE} 
& \textbf{R\textsuperscript{2}} \\
\hline
Linear Regression 
    & $0.890 \pm 0.017$ & $0.737 \pm 0.013$ 
    & $0.891 \pm 0.071$ & $0.721 \pm 0.070$ \\
Decision Tree 
    & $0.859 \pm 0.030$ & $0.755 \pm 0.016$ 
    & $1.128 \pm 0.104$ & $0.549 \pm 0.131$ \\
Random Forest 
    & $0.888 \pm 0.020$ & $0.738 \pm 0.015$ 
    & $0.890 \pm 0.069$ & $0.721 \pm 0.071$ \\
\hline
\end{tabular}
\end{table}

\begin{figure}[ht]
    \centering
    \begin{subfigure}[b]{0.48\linewidth}
        \centering
        \includegraphics[width=\linewidth]{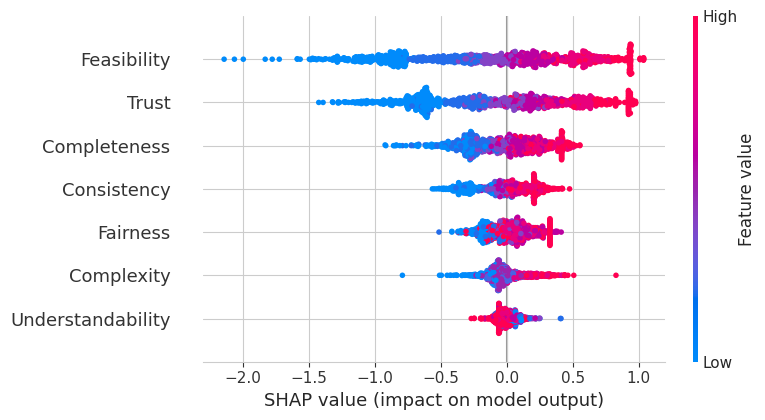}
        \caption{Global feature importance. Features are ranked by their mean absolute SHAP values, indicating their average impact on the predicted satisfaction.}
        \label{fig:shap_global}
    \end{subfigure}
    \hfill
    \begin{subfigure}[b]{0.48\linewidth}
        \centering
        \includegraphics[width=\linewidth]{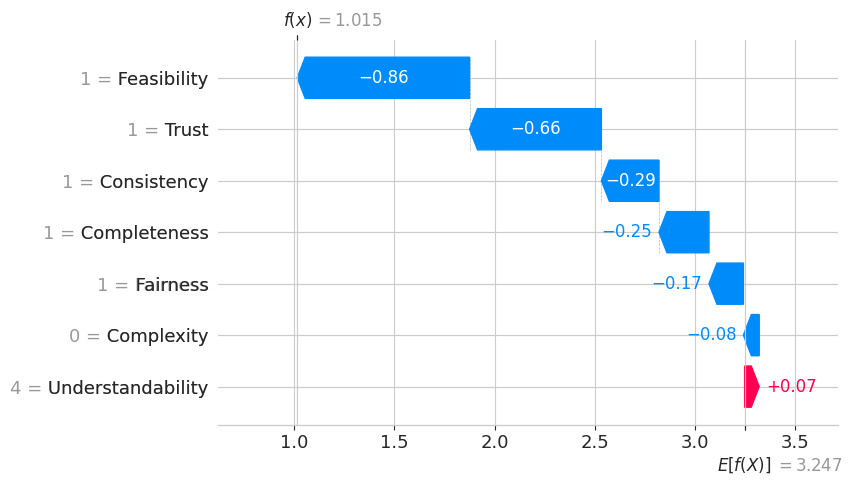}
        \caption{Local feature importance for Scenario 1. Red bars represent positive contributions increasing the predicted satisfaction, while blue bars indicate negative contributions.}
        \label{fig:shap_waterfall}
    \end{subfigure}
    \caption{SHAP analysis illustrating global and local feature importances for the Random Forest Regression model.}
    \label{fig:shap_regression}
\end{figure}

These results suggest that Trust and Feasibility stand out as the main predictors of Satisfaction in all models. The Decision Tree may capture slight non-linearities and interaction effects, achieving marginally better RMSE and $R^2$ than Linear Regression, while the Random Forest performance is comparable to linear regression on average, it lacks inherent interpretability. To clarify feature contributions in the Random Forest model, we applied SHAP (SHapley Additive exPlanations) analysis. SHAP values quantify the contribution of each feature to the predicted outcome. Figure~\ref{fig:shap_global} presents the global SHAP summary plot, confirming that Feasibility and Trust show the highest average impacts on predicted satisfaction, reinforcing findings from previous analyses.

 To illustrate how specific explanatory metrics influence individual predictions, we examined an instance (Scenario 1): \textit{"You are a 31-year-old divorced woman with high-school education, working 20 hours per week, earning below average salary. To earn above average, you must increase your education from high-school to Bachelor's degree."} Figure~\ref{fig:shap_waterfall} provides the SHAP waterfall plot for this scenario.

\subsection{Predicting Overall Satisfaction as a Classification Problem}
\begin{figure}[ht]
    \centering
    \begin{subfigure}[b]{0.32\linewidth}
        \centering
        \includegraphics[width=\linewidth]{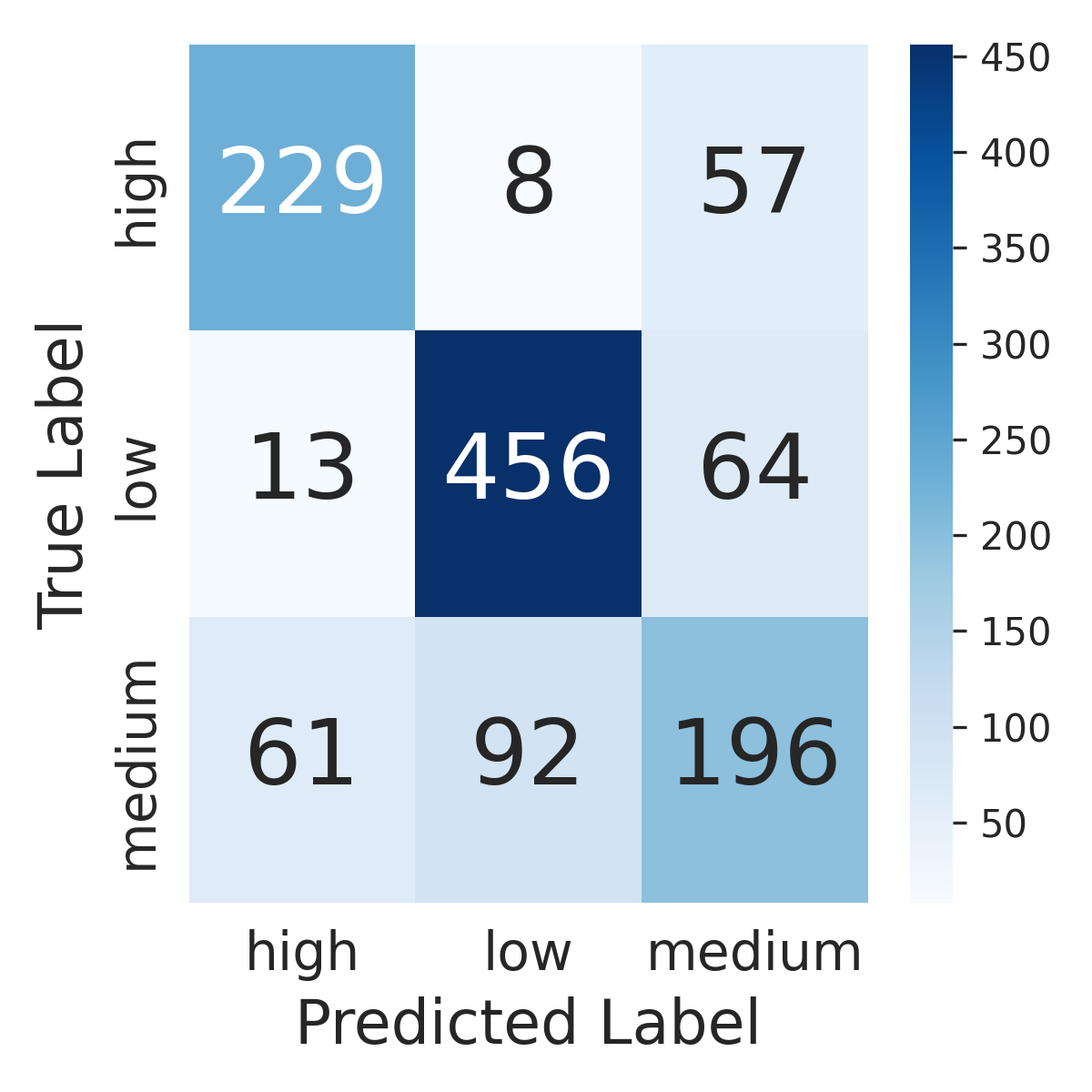}
        \caption{Logistic Regression}
        \label{fig:conf_log}
    \end{subfigure}
    \hfill
    \begin{subfigure}[b]{0.32\linewidth}
        \centering
        \includegraphics[width=\linewidth]{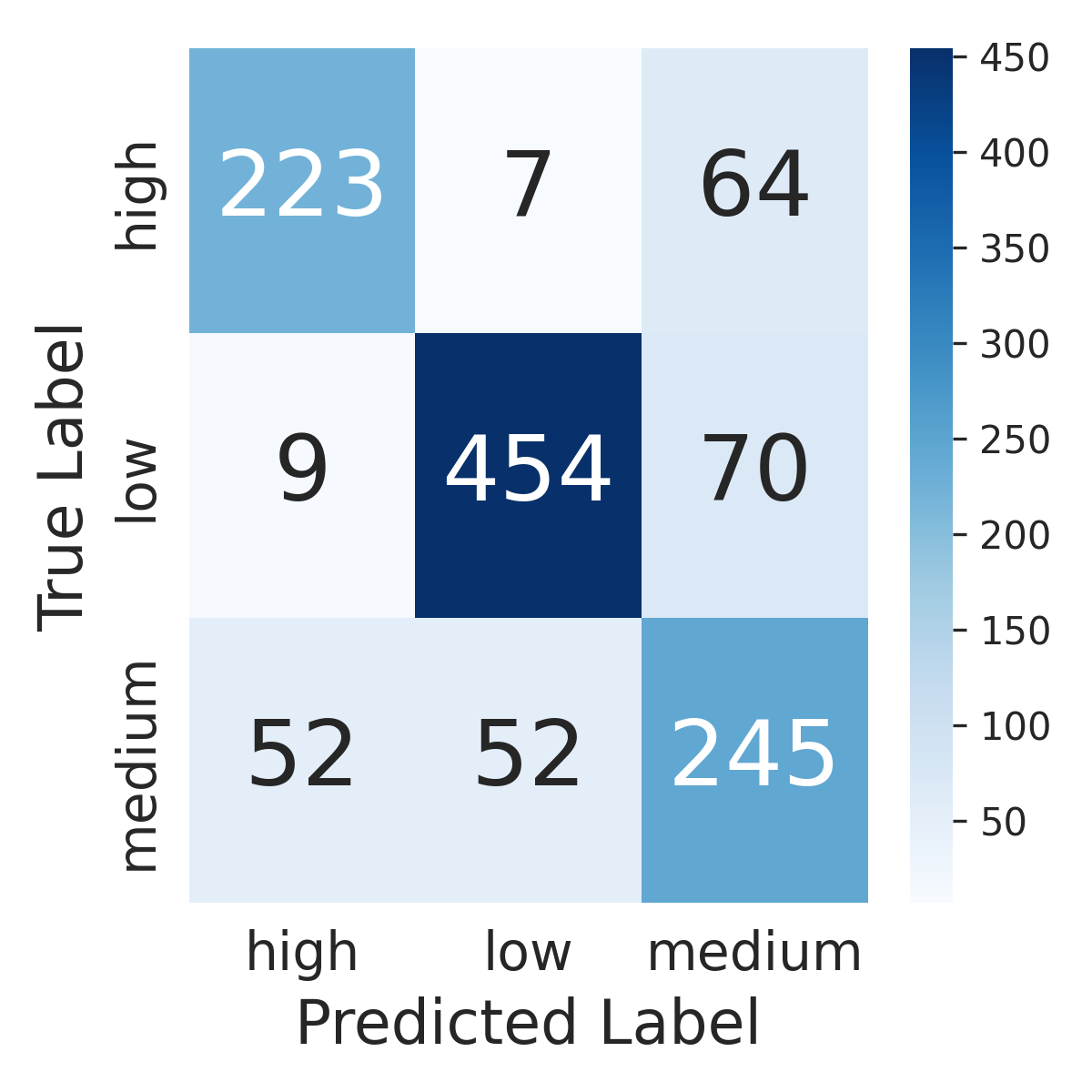}
        \caption{Decision Tree}
        \label{fig:conf_dec}
    \end{subfigure}
    \hfill
    \begin{subfigure}[b]{0.32\linewidth}
        \centering
        \includegraphics[width=\linewidth]{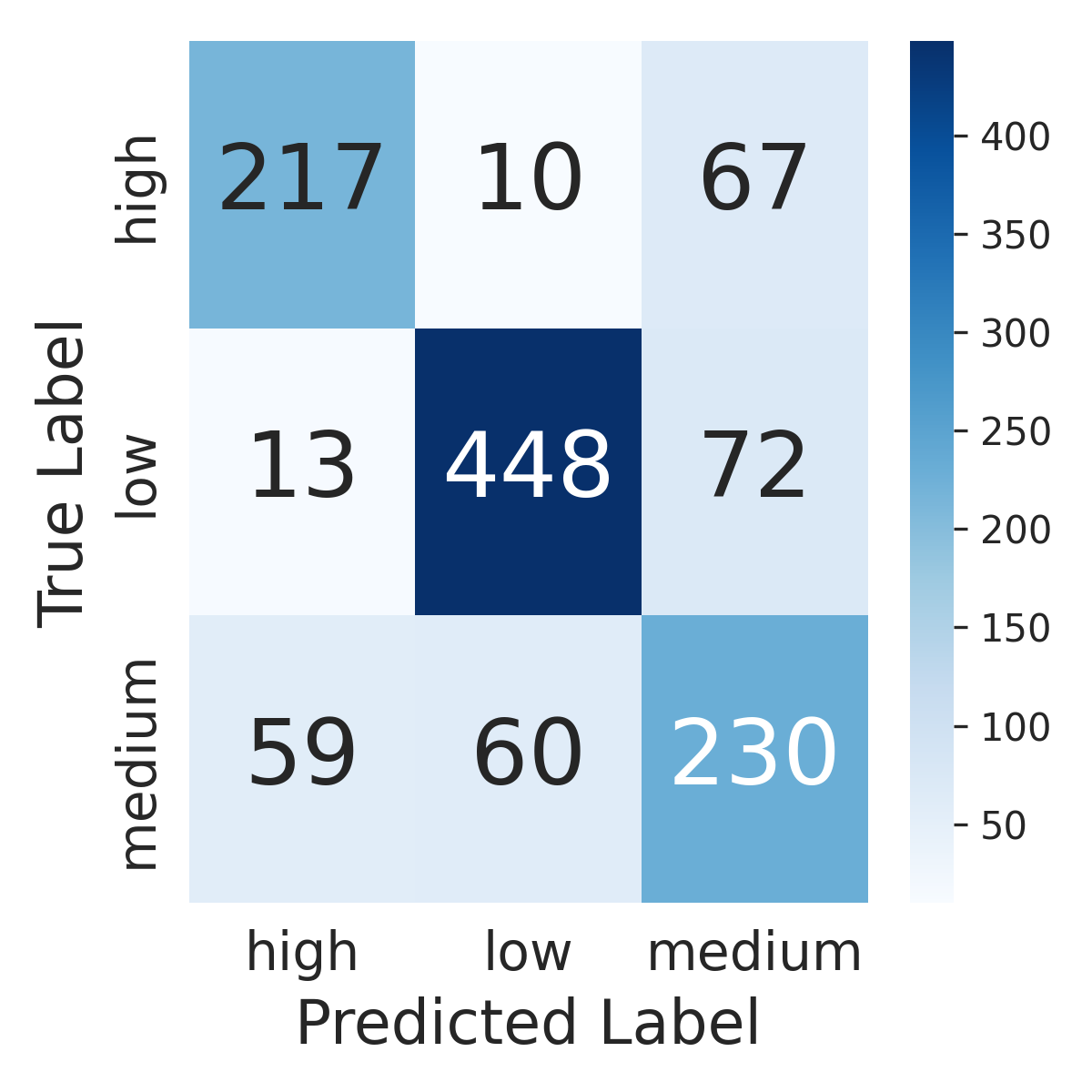}
        \caption{Random Forest}
        \label{fig:rand}
    \end{subfigure}
    \caption{Confusion matrices for the three models (Random Split).}
    \label{fig:confusion_matrices}
\end{figure}

\begin{figure}
\includegraphics[width=0.95\textwidth]{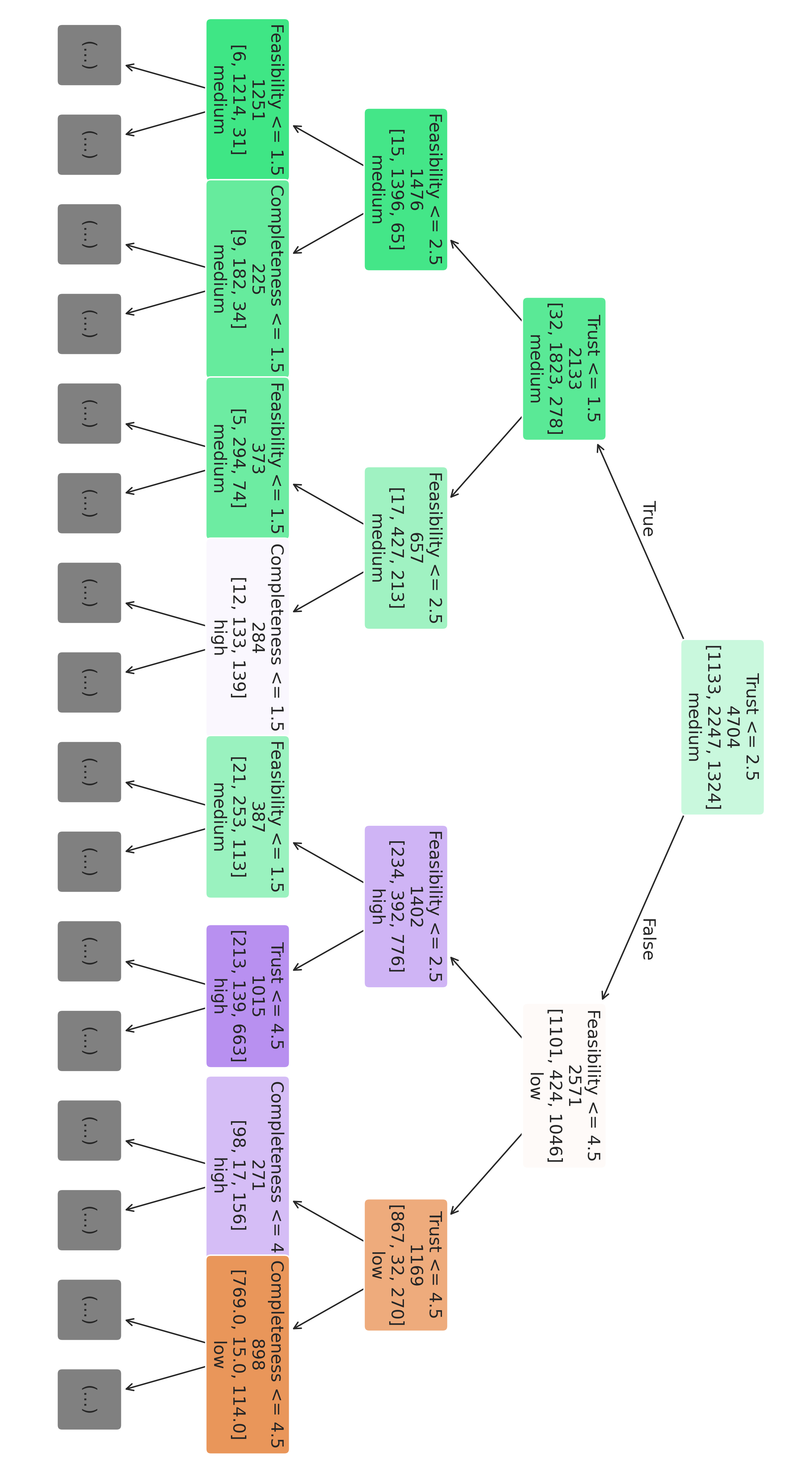}
\caption{\textbf{Decision Tree Trained on Scenario-Based Split.}
This tree classifies user satisfaction (low, medium, high) with seven explanation metrics. Node colors indicate the predicted class, and thresholds (e.g., \texttt{Trust} $\leq 2.5$) reflect how the model discriminates different satisfaction levels. } 
\label{fig:decision_tree}
\end{figure}

To investigate discrete satisfaction levels, we converted Overall Satisfaction into three classes: Low (1-2), Medium (3-4), High (5-6) with (2666, 1747, and 1467 instances, respectively). This led to a multi-class classification task. We applied Logistic Regression, Decision Tree, and Random Forest and similarly evaluated with 5-fold cross-validation under both random and scenario-based splits. To balance class distribution we used SMOTE-TOMEK resampling.
Figure~\ref{fig:confusion_matrices} shows the confusion matrices for each classifier. Logistic Regression (Figure~\ref{fig:conf_log}) tends to misclassify borderline cases of Medium vs. High, though it shows relatively fewer errors distinguishing Low from Medium. Decision Tree (Figure~\ref{fig:conf_dec}) overall has more correct classifications on the diagonal, particularly in distinguishing High from Medium, reflecting its higher accuracy in Table~\ref{tab:classification_comparison_2splits}. However, it can still confuse Medium with Low in certain scenarios. Random Forest (Figure~\ref{fig:rand}) exhibits a pattern similar to Logistic Regression, with Medium and High sometimes misclassified. Importantly, the Decision Tree allows for a transparent visualization of how metrics form top‐level splits when predicting satisfaction (Apendix A. Figure~\ref{fig:decision_tree}). The tree classifies user satisfaction (low, medium, high) by first splitting on 
\emph{Trust} and \emph{Feasibility}, with deeper nodes considering metrics like \emph{Completeness} or \emph{Complexity}.

Table~\ref{tab:classification_comparison_2splits} summarizes the mean accuracy and F1-scores (Macro) across 5 cross validation folds with the Decision Tree outperforming in the random split (highest accuracy/F1 of $\approx 0.78/0.77$) but being more sensitive to new scenarios (dropping to $\approx 0.70/0.68$). Logistic Regression and Random Forest demonstrate comparable performance, around $0.75$--$0.76$ accuracy, and they also more robust when tested for new, previously unseen explanations. 

\begin{table}[h!]
\centering
\caption{Comparison of classification model performances in predicting satisfaction classes using 5-fold cross-validation for two data-splitting strategies: random split and scenario-based split. Mean $\pm$ standard deviation are shown for both Accuracy and Macro F1.}
\label{tab:classification_comparison_2splits}
\begin{tabular}{|l|cc|cc|}
\hline
\multirow{2}{*}{\textbf{Model}}
  & \multicolumn{2}{c|}{\textbf{Random Split}}
  & \multicolumn{2}{c|}{\textbf{Scenario-Based Split}} \\
\cline{2-5}
  & \textbf{Accuracy} & \textbf{F1-score}
  & \textbf{Accuracy} & \textbf{F1-score } \\
\hline
Logistic Regression 
  & $0.757 \pm 0.008$ & $0.742 \pm 0.006$
  & $0.754 \pm 0.043$ & $0.734 \pm 0.029$ \\
Decision Tree
  & $0.781 \pm 0.009$ & $0.769 \pm 0.009$
  & $0.704 \pm 0.047$ & $0.680 \pm 0.036$ \\
Random Forest
  & $0.756 \pm 0.006$ & $0.739 \pm 0.004$
  & $0.756 \pm 0.046$ & $0.735 \pm 0.034$ \\
\hline
\end{tabular}
\end{table}

\begin{figure}[ht]
    \centering
    \begin{subfigure}[b]{0.32\linewidth}
        \centering
        \includegraphics[width=\linewidth]{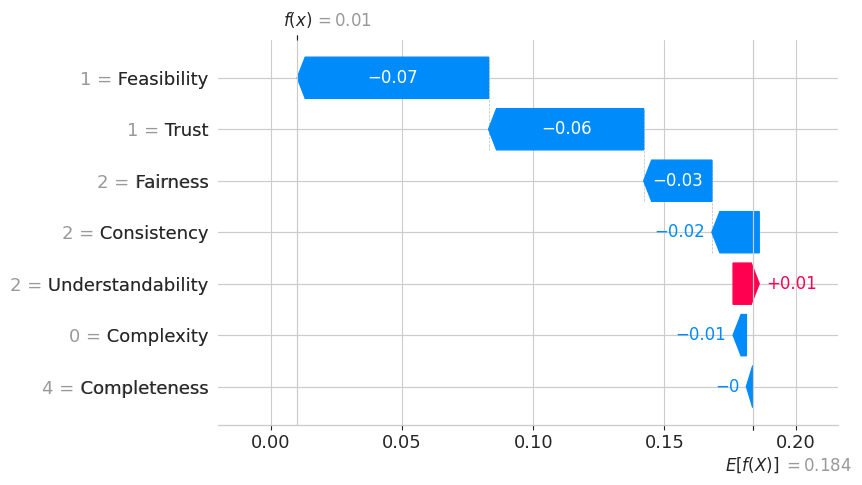}
        \caption{'low' satisfaction}
        \label{fig:shap_low}
    \end{subfigure}
    \hfill
    \begin{subfigure}[b]{0.32\linewidth}
        \centering
        \includegraphics[width=\linewidth]{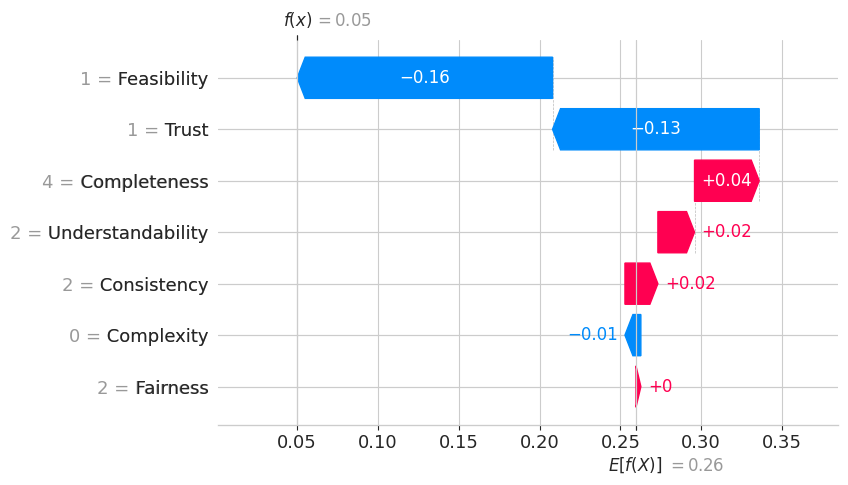}
        \caption{'medium' satisfaction}
        \label{fig:shap_mid}
    \end{subfigure}
    \hfill
    \begin{subfigure}[b]{0.32\linewidth}
        \centering
        \includegraphics[width=\linewidth]{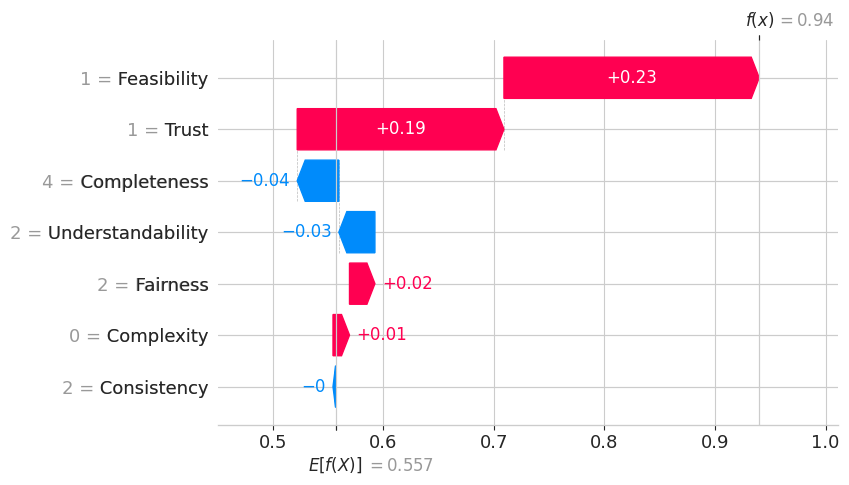}
        \caption{'high' satisfaction}
        \label{fig:shap_high}
    \end{subfigure}
    \caption{SHAP waterfall plots illustrating feature contributions for each predicted satisfaction class for Scenario 1: \textit{"You are a 31-year-old divorced woman with high-school education, working 20 hours per week, earning below average salary. To earn above average, you must increase your education from high-school to Bachelor's degree."}}
    \label{fig:shap_classification_individual}
\end{figure}

Similarly, we conducted SHAP analysis of local feature contributions for different satisfaction classes (low, medium, high) for Scenario 1 using the Random Forest classification model  (Fig.~\ref{fig:shap_classification_individual}). Consistent with previous findings, Feasibility and Trust are the dominant predictors across all classes. However, other metrics such as Completeness, Fairness and Consistency are decisive factors differentiating between low and medium satisfaction classes.

Of particular note, the scenario-based split shows higher variation during cross-validation, reflecting the diverse complexity of different counterfactual scenarios. Because certain scenarios target specific biases and present greater predictive challenges than others, some scenario subsets are inherently easier or harder to model, resulting in more variable performance. This performance drop highlights the challenge of generalizing predictions to entirely new scenarios, underscoring the importance of robust scenario diversity in training data.

\section{Discussion}
This study set out to model overall satisfaction with counterfactual explanations by examining user ratings on multiple explanatory metrics (Feasibility, Consistency, Completeness, Trust, Fairness, Complexity, and Understandability). The results consistently highlight Feasibility and Trust as the most influential factors in predicting user satisfaction. \textit{Feasibility}, defined as the degree to which recommended changes are realistic and actionable, and \textit{Trust}, understood as the belief that the suggested changes would bring about the intended outcome, both stood out through large coefficients in linear regression and prominent splits in decision trees. Their dominance confirms longstanding user-centric observations in explainable AI research: participants place high value on seeing counterfactuals that they can readily implement and that they can credibly believe will lead to the desired effect. Importantly, \textit{Completeness} also emerged as a contributor in some of our analyses, though it did not overshadow the two primary drivers of satisfaction, suggesting that in certain domains users favor more detailed explanations if those explanations are simultaneously feasible and trustworthy. At the same time, our analyses show that even when Feasibility and Trust are omitted, the remaining metrics (Completeness, Consistency, Fairness, Complexity, Understandability) still capture around 58\% of the variance in satisfaction. This suggests that \emph{other explanatory qualities also remain relevant in shaping overall satisfaction, though Feasibility and Trust are clearly the most influential}. 

One central methodological insight is that \emph{asking users about multiple explanatory metrics, rather than only about overall satisfaction, brings more nuanced understanding of why an explanation succeeds or fails}. Our correlation and factor analyses revealed strong overlap among many of these metrics—Feasibility, Trust, Completeness, Consistency, and Fairness loaded together on a primary factor, while Complexity appeared partly independent. It indicates that participants generally perceive a holistic notion of a “good” or “poor” explanation. This pattern implies that, for many users, evaluating a counterfactual is not a matter of distinctly disentangling fairness, consistency, or completeness, but rather forming a more integrated sense of overall quality. At the same time, Complexity appears to be at least partly independent, which aligns with the idea that \emph{an explanation can be somewhat complex yet still garner high satisfaction, provided it remains actionable and believable}. 

By comparison, Fairness did not emerge as driver of satisfaction in this dataset. Still, every scenario with low fairness also ranked with low Satisfaction, suggesting that fairness’s impact may have been absorbed by correlated metrics. In real applications, fairness can be critical for decisions involving potential discrimination (e.g., loan approvals, medical treatments), while understandability might become vital if the system outputs specialized or highly technical explanations \cite{goethals2024precof}. Future studies should examine user groups that are more sensitive to these specific metrics. One of the findings was that human evaluation is is significantly influenced by user backgrounds (medical professionals and machine learning experts). Individuals with a strong clinical focus might value feasibility and clarity differently than those trained in algorithmic reasoning, while ML experts could emphasize completeness or consistency more strongly \cite{ghassemi2021false,bansal2019beyond,10.1145/3613904.3642474}.

In practical terms, these findings underscore several implications for counterfactual explanation algorithms design. First, the emphasis on Feasibility suggests that XAI systems should ensure their generated counterfactuals appear realistic to end users, for instance by embedding domain constraints or avoiding implausible feature changes. Second, the prominence of Trust implies that systems should provide evidence of the proposed modifications’ effects (e.g., model confidence or analogy to past successes). Third, while Completeness is secondary, there may be domains where users do want more detailed explanations covering multiple angles of the model’s logic.

\subsubsection{Limitations and Future Work}
The 30 scenarios in our dataset, while varied, do not represent the full spectrum of real-world counterfactuals, especially regarding fairness or domain-specific constraints. Nevertheless, it is important to note that average time of evaluating those scenarios is 42 minutes. Future work should aim at gathering similar dataset but for specialized application and with different levels of expert knowledge to generalize modeling satisfaction question.  The participant group of roughly 200 respondents is fairly large for a user study, yet it may not reflect domain experts or culturally varied perspectives, more specialized user group might value fairness or completeness more strongly \cite{goethals2024precof}. There are inherent limitations in online studies, such as the inability to ensure that participants completed the questionnaire without distractions. Also, we did not systematically vary fairness or clarity, their roles might be underrepresented here. Future work should examine more diverse or extreme variations in these factors to see if they become decisive under particular conditions.

Despite these limitations, our findings provide a tangible contribution to understanding overall satisfaction through the lens of key explanatory qualities. We show that focusing on a handful of core metrics, especially feasibility, trustworthiness, and completeness, goes a long way toward predicting user contentment with counterfactual explanations. At the same time, complexity appears relatively independent, suggesting that adding details or allowing higher complexity does not need to undermine satisfaction if the explanation remains realistically implementable and fosters confidence in its correctness. This insight can guide the dilemma of explaining 'less is more' versus 'fully detailed is better'. Future research could validate these findings in other domains (e.g., medical, financial) and with specialized user cohorts (e.g., clinicians, ML engineers). By refining how we measure and optimize these explanatory metrics for diverse audiences, we can further advance the practical and ethical utility of explainable AI systems.

\subsubsection{Code Availability}
The implementation of the models and analyses presented in this paper is publicly available at \href{https://github.com/anitera/CounterEval/tree/main/predicting_satisfaction}{CounterEval GitHub repository}.
%
%

\begin{credits}
\subsubsection{\ackname} This research was supported by Estonian Research Council Grants PRG1604 and PUTJD1252, the European Union’s Horizon 2020 Research and Innovation Programme under Grant Agreement No. 952060 (Trust AI), Foundations of Secure Digital Solutions and Artificial Intelligence TEM-TA119, the Estonian Centre of Excellence in Artificial Intelligence (EXAI), funded by the Estonian Ministry of Education and Research grant TK213.

\subsubsection{\discintname}
The authors have no competing interests to declare that are
relevant to the content of this article.
\end{credits}
%
%
%
\bibliographystyle{splncs04}
\bibliography{bibliography}

\begin{thebibliography}{10}
\providecommand{\url}[1]{\texttt{#1}}
\providecommand{\urlprefix}{URL }
\providecommand{\doi}[1]{https://doi.org/#1}

\bibitem{adadi2018peeking}
Adadi, A., Berrada, M.: Peeking inside the black-box: a survey on explainable artificial intelligence (xai). IEEE access  \textbf{6},  52138--52160 (2018)

\bibitem{bansal2019beyond}
Bansal, G., Nushi, B., Kamar, E., Lasecki, W.S., Weld, D.S., Horvitz, E.: Beyond accuracy: The role of mental models in human-ai team performance. In: Proceedings of the AAAI conference on human computation and crowdsourcing. vol.~7, pp. 2--11 (2019)

\bibitem{barbu2024exploring}
Barbu, E., Domnich, M., Vicente, R., Sakkas, N., Morim, A.: Exploring commonalities in explanation frameworks: A multi-domain survey analysis. arXiv preprint arXiv:2405.11958  (2024)

\bibitem{Byrne2019}
Byrne, R.M.J.: Counterfactuals in explainable artificial intelligence (xai): Evidence from human reasoning. In: Proceedings of the Twenty-Eighth International Joint Conference on Artificial Intelligence, {IJCAI-19}. pp. 6276--6282. International Joint Conferences on Artificial Intelligence Organization (7 2019). \doi{10.24963/ijcai.2019/876}

\bibitem{byrne2005rational}
Byrne, R.M.: The rational imagination (2005)

\bibitem{de2024evaluating}
De~Bona, F.B., Dominici, G., Miller, T., Langheinrich, M., Gjoreski, M.: Evaluating explanations through llms: Beyond traditional user studies. arXiv preprint arXiv:2410.17781  (2024)

\bibitem{domnich2024towards}
Domnich, M., Valja, J., Veski, R.M., Magnifico, G., Tulver, K., Barbu, E., Vicente, R.: Towards unifying evaluation of counterfactual explanations: Leveraging large language models for human-centric assessments. arXiv preprint arXiv:2410.21131  (2024)

\bibitem{codice_domnich_2024}
Domnich, M., Vicente, R.: Enhancing counterfactual explanation search with diffusion distance and directional coherence. In: Longo, L., Lapuschkin, S., Seifert, C. (eds.) Explainable Artificial Intelligence. pp. 60--84. Springer Nature Switzerland, Cham (2024)

\bibitem{domnich_2024_14672264}
Domnich, M., Välja, J., Veski, R.M., Magnifico, G., Tulver, K., Barbu, E., Vicente, R.: Countereval: Towards unifying evaluation of counterfactual explanations (Dec 2024). \doi{10.57967/hf/3824}, \url{https://doi.org/10.57967/hf/3824}

\bibitem{10.1145/3613904.3642474}
Ehsan, U., Passi, S., Liao, Q.V., Chan, L., Lee, I.H., Muller, M., Riedl, M.O.: The who in xai: How ai background shapes perceptions of ai explanations. In: Proceedings of the 2024 CHI Conference on Human Factors in Computing Systems. CHI '24, Association for Computing Machinery, New York, NY, USA (2024). \doi{10.1145/3613904.3642474}, \url{https://doi.org/10.1145/3613904.3642474}

\bibitem{forster2021capturing}
Förster, M., Hühn, P., Klier, M., Kluge, K.: Capturing users' reality: A novel approach to generate coherent counterfactual explanations. In: 54th Hawaii International Conference on System Sciences, HICSS 2021, Kauai, Hawaii, USA, January 5, 2021. pp. 1--10. ScholarSpace (2021), \url{http://hdl.handle.net/10125/70767}

\bibitem{ghassemi2021false}
Ghassemi, M., Oakden-Rayner, L., Beam, A.L.: The false hope of current approaches to explainable artificial intelligence in health care. The Lancet Digital Health  \textbf{3}(11),  e745--e750 (2021)

\bibitem{goethals2024precof}
Goethals, S., Martens, D., Calders, T.: Precof: counterfactual explanations for fairness. Machine Learning  \textbf{113}(5),  3111--3142 (2024)

\bibitem{guidotti_counterfactual_2022}
Guidotti, R.: Counterfactual explanations and how to find them: literature review and benchmarking. Data Mining and Knowledge Discovery  (2022). \doi{10.1007/s10618-022-00831-6}, \url{https://doi.org/10.1007/s10618-022-00831-6}

\bibitem{karimi2022survey}
Karimi, A.H., Barthe, G., Sch{\"o}lkopf, B., Valera, I.: A survey of algorithmic recourse: contrastive explanations and consequential recommendations. ACM Computing Surveys  \textbf{55}(5),  1--29 (2022)

\bibitem{keane2021bettercounterfactualexplanationskey}
Keane, M.T., Kenny, E.M., Delaney, E., Smyth, B.: If only we had better counterfactual explanations: Five key deficits to rectify in the evaluation of counterfactual xai techniques (2021), \url{https://arxiv.org/abs/2103.01035}

\bibitem{miller_explanation_2019}
Miller, T.: Explanation in artificial intelligence: Insights from the social sciences. Artificial Intelligence  \textbf{267},  1--38 (2019). \doi{10.1016/j.artint.2018.07.007}, \url{https://www.sciencedirect.com/science/article/pii/S0004370218305988}

\bibitem{mothilal_explaining_2020}
Mothilal, R.K., Sharma, A., Tan, C.: Explaining machine learning classifiers through diverse counterfactual explanations. In: Proceedings of the 2020 Conference on Fairness, Accountability, and Transparency. pp. 607--617. {FAT}* '20, Association for Computing Machinery (2020). \doi{10.1145/3351095.3372850}, \url{https://doi.org/10.1145/3351095.3372850}, event-place: Barcelona, Spain

\bibitem{rasouli2024care}
Rasouli, P., Chieh~Yu, I.: Care: Coherent actionable recourse based on sound counterfactual explanations. International Journal of Data Science and Analytics  \textbf{17}(1),  13--38 (2024)

\bibitem{van2021interpretable}
Van~Looveren, A., Klaise, J.: Interpretable counterfactual explanations guided by prototypes. In: Joint European Conference on Machine Learning and Knowledge Discovery in Databases. pp. 650--665. Springer (2021)

\bibitem{vannostrand2024actionable}
VanNostrand, P.M., Hofmann, D.M., Ma, L., Rundensteiner, E.A.: Actionable recourse for automated decisions: Examining the effects of counterfactual explanation type and presentation on lay user understanding. In: Proceedings of the 2024 ACM Conference on Fairness, Accountability, and Transparency. pp. 1682--1700 (2024)

\bibitem{wachter2017counterfactual}
Wachter, S., Mittelstadt, B., Russell, C.: Counterfactual explanations without opening the black box: Automated decisions and the gdpr. Harv. JL \& Tech.  \textbf{31}, ~841 (2017)

\bibitem{wang2021explanations}
Wang, X., Yin, M.: Are explanations helpful? a comparative study of the effects of explanations in ai-assisted decision-making. In: Proceedings of the 26th International Conference on Intelligent User Interfaces. pp. 318--328 (2021)

\bibitem{warren2023categorical}
Warren, G., Byrne, R.M.J., Keane, M.T.: Categorical and continuous features in counterfactual explanations of ai systems. In: Proceedings of the 28th International Conference on Intelligent User Interfaces. p. 171–187. IUI '23, Association for Computing Machinery, New York, NY, USA (2023). \doi{10.1145/3581641.3584090}, \url{https://doi.org/10.1145/3581641.3584090}

\bibitem{zemla2017evaluating}
Zemla, J.C., Sloman, S., Bechlivanidis, C., Lagnado, D.A.: Evaluating everyday explanations. Psychonomic bulletin \& review  \textbf{24},  1488--1500 (2017)

\end{thebibliography}

\section*{A Appendix}

\end{document}